\documentclass{aa}  
\usepackage{natbib}
\bibpunct{(}{)}{,}{a}{}{,} % to follow the A&A style
\usepackage{amsmath}
\usepackage{graphicx}
\usepackage{longtable}
\usepackage{txfonts}

\begin{document}
\newcommand{\kmps}{km~s$^{-1}$}
\newcommand{\kmpsb}{km~s$^{-1}$ }
\newcommand{\cdho}{C$^{18}$O}
\newcommand{\cdhob}{C$^{18}$O }
\newcommand{\cdso}{C$^{17}$O}
\newcommand{\cdsob}{C$^{17}$O }
\newcommand{\dzco}{$^{12}$CO}
\newcommand{\dzcob}{$^{12}$CO }
\newcommand{\tzco}{$^{13}$CO}
\newcommand{\tzcob}{$^{13}$CO }
\newcommand{\ndhp}{N$_2$H$^+$}
\newcommand{\ndhpb}{N$_2$H$^+$ }
\newcommand{\nddp}{N$_2$D$^+$}
\newcommand{\nddpb}{N$_2$D$^+$ }
\newcommand{\hdb}{H$_2$ }
\newcommand{\ddb}{D$_2$ }
\newcommand{\htp}{H$_3^+$}
\newcommand{\htpb}{H$_3^+$ }
\newcommand{\dtp}{D$_3^+$}
\newcommand{\dtpb}{D$_3^+$ }
\newcommand{\hddp}{H$_2$D$^+$}
\newcommand{\hddpb}{H$_2$D$^+$ }
\newcommand{\ddhp}{D$_2$H$^+$}
\newcommand{\ddhpb}{D$_2$H$^+$ }
\newcommand{\cd}{column density}
\newcommand{\cdb}{column density }
\newcommand{\cc}{cm$^{-3}$}
\newcommand{\ccb}{cm$^{-3}$ }
\newcommand{\sqc}{cm$^{-2}$}
\newcommand{\sqcb}{cm$^{-2}$ }
\newcommand{\ctds}{C$^{32}$S}
\newcommand{\ctdsb}{C$^{32}$S }
\newcommand{\ctqs}{C$^{34}$S}
\newcommand{\ctqsb}{C$^{34}$S }
\newcommand{\tdso}{$^{32}$SO}
\newcommand{\tdsob}{$^{32}$SO }
\newcommand{\tqso}{$^{34}$SO}
\newcommand{\tqsob}{$^{34}$SO }
\newcommand{\juz}{(J:1--0)}
\newcommand{\juzb}{(J:1--0) }
\newcommand{\jdu}{(J:2--1)}
\newcommand{\jdub}{(J:2--1) }
\newcommand{\jtd}{(J:3--2)}
\newcommand{\jtdb}{(J:3--2) }
\newcommand{\jqt}{(J:4--3)}
\newcommand{\jqtb}{(J:4--3) }
\newcommand{\jkk}{J$_\mathrm{KK\arcsec}$}
\newcommand{\jkkb}{J$_\mathrm{KK\arcsec}$ }
\newcommand{\nhdd}{NH$_2$D}
\newcommand{\nhddb}{NH$_2$D }
\newcommand{\den}{n(H$_2$)}
\newcommand{\denb}{n(H$_2$) }
\newcommand{\mjy}{MJy/sr}%$^{-1}$}
\newcommand{\mjyb}{MJy/sr}%$^{-1}$ }
\newcommand{\Av}{A$_{\mathrm V}$}
\newcommand{\Avb}{A$_{\mathrm V}$ }
\newcommand{\SM}{M$_\odot$}
\newcommand{\SMb}{M$_\odot$ }
\newcommand{\pdix}[1]{$\times$~10$^{#1}$}
\newcommand{\pdixb}[1]{$\times$~10$^{#1}$ }

   \title{Chemical modeling of \object{L183} (= \object{L134N})\,: an estimate of the ortho/para H$_2$ ratio}
   %The ortho/para ratio of H$_2$ as a probe of the dense regions of the \object{L183} (=\object{L134N}) complex}
%   \title{\hddpb as a probe of the dense regions of the \object{L183} (=\object{L134N}) complex}

%   \subtitle{I. Estimates of ortho/para \hddpb from \n2dp/\n2hp}

   \author{L. Pagani           \inst{1}
          \and
          C. Vastel\inst{2}
          \and
          E. Hugo\inst{3}
          \and
          V. Kokoouline\inst{4}
          \and
	Chris H. Greene\inst{5}
          \and
           A. Bacmann\inst{6}
          \and
         E. Bayet\inst{7}
          \and
          C. Ceccarelli\inst{6}
          \and
	R. Peng\inst{8}
          \and
         S. Schlemmer\inst{3}
         }

   \offprints{L.Pagani}

 \institute{ LERMA \& UMR8112 du CNRS, Observatoire de
  Paris, 61, Av. de l'Observatoire, 75014 Paris, France\\
\email{laurent.pagani@obspm.fr}
\and
CESR, 9 avenue du colonel Roche, BP44348, Toulouse Cedex 4, France\\
 \email{vastel@cesr.fr}
\and
I. Physikalisches Institut,
Universit\"at zu K\"oln,
Z\"ulpicher Strasse 77,
50937 K\"oln,
Germany
\\
 \email{hugo@ph1.uni-koeln.de,schlemmer@ph1.uni-koeln.de}
 \and
Department of Physics, University of Central Florida, Orlando, FL-32816, USA\\
 \email{slavako@physics.ucf.edu}
 \and
Department of Physics and JILA, University of Colorado, Boulder, Colorado 80309-0440, USA\\
 \email{chris.greene@colorado.edu}
\and
Laboratoire d'Astrophysique de Grenoble,
  Universit\'e Joseph Fourier, UMR5571 du CNRS, 414 rue de la Piscine,
  38041 Grenoble Cedex 09, France\\
  \email{cecilia.ceccarelli@obs.ujf-grenoble.fr,aurore.bacmann@obs.ujf-grenoble.fr}
\and
 Department of physics and astronomy, University College London, Gower street, London WC1E 6BT, UK\\
 \email{eb@star.ucl.ac.uk}
 \and
 Caltech Submillimeter Observatory, 111 Nowelo Street, Hilo, HI 96720, USA\\
 \email{peng@submm.caltech.edu}
           }

   \date{Received 15/07/2008; accepted \today}

% \abstract{}{}{}{}{} 
% 5 {} token are mandatory
 
  \abstract
   {The high degree of deuteration observed in some prestellar cores depends on the ortho-to-para H$_2$ ratio through the \htpb fractionation.}
   {We want to constrain the ortho/para H$_2$ ratio across the L183 prestellar core. This is required to correctly describe the deuteration amplification phenomenon in depleted cores such as L183 and to relate the total (ortho+para) \hddpb abundance to the sole ortho--\hddpb column density measurement.}
   {To constrain this ortho/para H$_2$ ratio and derive its profile, we make use of the N$_2$D$^+$/N$_2$H$^+$ ratio and of the ortho--\hddp observations performed across the prestellar core. We use two simple chemical models limited to an almost totally depleted core description. New dissociative recombination and trihydrogen cation--dihydrogen reaction rates (including all isotopologues) are presented in this paper and included in our models.}
    {We estimate the H$_2$D$^+$ ortho/para ratio in the L183 cloud, and constrain the H$_2$ ortho/para ratio\,: we show that it varies across the prestellar core by at least an order of magnitude, being still very high ($\approx$ 0.1) in most of the cloud. Our  time-dependent model indicates that the prestellar core is presumably older than 1.5--2 \pdix{5} years but that it may not be much older. We also show that it has reached its present density only recently and that its contraction from a uniform density cloud can be constrained.}
   {A proper understanding of deuteration chemistry cannot be attained without taking into account the whole ortho/para family of molecular hydrogen and trihydrogen cation isotopologues as their relations are of utmost importance in the global scheme. Tracing the ortho/para H$_2$ ratio should also place useful constraints on the dynamical evolution of prestellar cores.}


   \maketitle
%
%________________________________________________________________

\section{Introduction}

Studies of the earliest stages of star formation have increased in the last ten years with the advent of new receivers acquiring 
better spatial and spectral resolution. Prestellar cores are dense and cold cores where gravitational collapse has not yet occured. 
In the densest regions of the core (where n$_{H_2}$ is greater than a few 10$^4$ cm$^{-3}$) most heavy molecules containing carbon, nitrogen and 
oxygen seem to deplete onto the dust grains  and only light ions remain in the gas phase. There has been extensive observational evidence of CO 
and CS depletion in the center of prestellar cores \citep[e.g.][]{Caselli99,Bergin02, Bacmann02,Tafalla04,Pagani05} which seems to be typical of the majority 
of dense cores. Nitrogen--bearing species like CN, NH$_3$ and N$_2$H$^+$ appear 
to subsist longer before freezing-out onto the dust grains \citep{Tafalla06,Pagani07,Hily08}. \\
In conditions under which heavy species are depleted, H$^+$ and \htpb (and its deuterated counterparts) are the most abundant ions 
subsisting in the gas phase. \hddpb has been widely detected and mapped in protostars and prestellar cores \citep{Caselli03,Vastel06,Caselli08} through its ortho 
fundamental line, with abundances high enough to be explained by the CO depletion theory/observations. Although 
difficult to observe from Earth, the \ddhpb molecule has been detected with its para line towards the 16293E prestellar core in the L1689N molecular cloud \citep{Vastel04} with an abundance similar 
to the ortho--\hddp molecule as suggested by \citet{Phillips03}. \\
Consequently many theoretical studies were performed that included all the deuterated forms of the \htpb ion \citep[e.g.][]{Roberts03,Walmsley04}. 
However all nuclear spin states (ortho, meta, para, corresponding to the spin state of the protons or deuterons) must be taken into account in order 
to compare with the observational sets. Moreover the thermicity of the forward/backward reactions strongly depends on the symmetric state of the 
species. Though the influence of the ortho/para H$_2$ ratio on the chemistry of \hddpb \citep{Pagani92} had been described several years before 
the first detection of the ion \citep{Stark99}, it is only recently that this specific spin-state dependent chemistry has been studied in detail \citep{Flower04, Flower06a}. \\ 

The motivation for our study is the many deuterated observations performed in the L183 prestellar core (PSC) and the main aim of this paper is to study the ortho/para H$_2$ ratio from some of these deuterated species observations. To this effect, we constrain two 
chemical models including all the symmetric states of H$_2$, D$_2$, \htpb and its deuterated counterparts with observations of ortho--\hddp, combined with previous \ndhpb and \nddpb observations. These models have been set up 
using recent dissociative recombination rates computed for \htpb and its isotopologues as well as all non negligible reaction rates between H$_2$ and \htpb and their isotopologues (both presented for the 
first time in this paper).

%__________________________________________________________________

\section{Observations}

\subsection{Deuterated \htp}

We first observed the ortho--\hddpb (and para--\ddhp) with the Caltech Submillimeter Observatory (CSO) monopixel receiver and subsequently took advantage of the newly installed Heterodyne ARray Program, 16 channel 350 GHz band ("B-band") (HARP-B) camera on the James Clerk Maxwell Telescope (JCMT) to fully map its emission.

\subsubsection{CSO observations}

The \hddpb and \ddhpb observations were carried out at the CSO, 
between August 2004 and April 2005, under good weather conditions (225 GHz zenith opacity 
always less than 0.06). Scans were taken, using the chopping secondary 
with a throw of 3$^{\prime}$, using the reference position: $\alpha_{2000}~=~15^h~54^m~08^s.50$, 
$\delta_{2000}~=~-02^\circ~52^\prime~48^{\prime \prime}$.

\begin{figure}
\centering
\includegraphics[height=10cm]{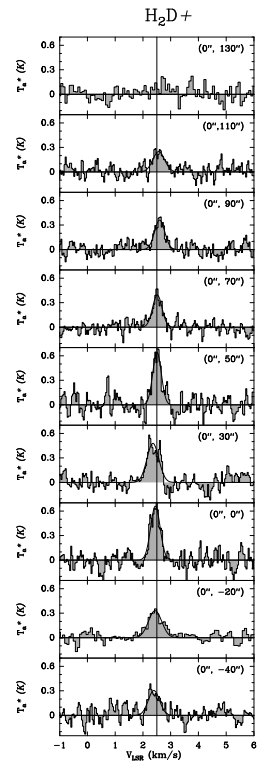}
\caption{CSO map of the \hddpb (1$_{10}$-1$_{11}$) line. The position is indicated in arcseconds for each spectrum and 
the (0,0) position corresponds to $\alpha_{2000}~=~15^h~54^m~08^s.50$, $\delta_{2000}~=~-02^\circ~52^\prime~48^{\prime \prime}$.
The Y-axis represents the antenna temperature}
\label{specmap}
\end{figure}

\begin{figure}
\centering
\includegraphics[height=8.5cm,angle=-90]{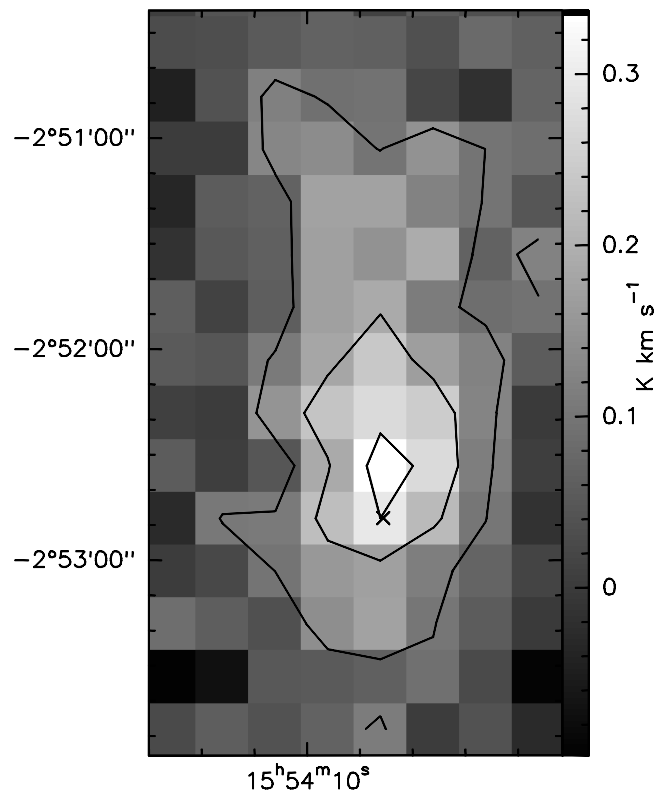}
\caption{JCMT map of the \hddpb (1$_{10}$-1$_{11}$) line. The dust peak position \citep{Pagani03} is indicated by a cross and corresponds to $\alpha_{2000}~=~15^h~54^m~08^s.50$, $\delta_{2000}~=~-02^\circ~52^\prime~48^{\prime \prime}$. Contour levels are 0.1, 0.2 and 0.3 K km s$^{-1}$}
\label{intensmap}
\end{figure}

\begin{table*}
\caption[]{Line parameters from the JCMT and CSO observations. The positions are offsets to the dust peak emission at 
$\alpha_{2000}~=~15^h~54^m~08^s.50$, $\delta_{2000}~=~-02^\circ~52^\prime~48^{\prime \prime}$. 
For non--detected lines we give the 3$\sigma$ upper limit. For JCMT and p--\ddhpb at CSO, we give both the Monte Carlo (MC) and the LTE column density estimates.} 
\label{jcmt_parameters}
\begin{tabular}{cccccccc}
\hline
                            &                                  &            &&       JCMT          &         &\\
\hline
Position                      &  rms           &   $\delta$v              &  $\Delta$v    &   T$_{a}^*$  & N(o--\hddp)$^\mathrm{a,b}_\mathrm{MC}$& N(o--\hddp)$^\mathrm{b}_\mathrm{LTE}$  & tau$^\mathrm{b}_\mathrm{LTE}$\\
($^{\prime\prime}$)  & (K)            & km~s$^{-1}$  & km~s$^{-1}$ & K                 &   cm$^{-2}$ &   cm$^{-2}$ & \\  
\hline
(-45,0)                    & 0.08           &  0.049       &   0.3      &   0.11  &  3.6 10$^{12}$&2.9 10$^{12}$    & 0.11\\
(-30,0)	                    & 0.08           &  0.049       &   0.26      &   0.28  &  8.1 10$^{12}$&8.1 10$^{12}$    & 0.31\\
(-15,0)	                     & 0.09           &  0.049       &   0.51      &   0.41  &  2.0 10$^{13}$&1.3 10$^{13}$      & 0.49\\
(0,0)                        & 0.09         &  0.049       &   0.41      &   0.57  & 2.3 10$^{13}$&2.0 10$^{13}$     & 0.77\\
(15,0)                      & 0.08          &  0.049       &   0.47      &   0.46  &   2.0 10$^{13}$&1.5 10$^{13}$  & 0.57\\
(30,0)                      & 0.08           &  0.049       &   0.39      &   0.21  & 8.1 10$^{12}$&5.7 10$^{12}$     &     0.22   \\
(45,0)                      & 0.08           &  0.049       &   0.50      &   0.13  &  3.6 10$^{12}$&3.4 10$^{12}$   & 0.13\\
(60,0)                      &    0.08       &  0.049       &        & $<$0.06    &   $<$1.8 10$^{12}$&$<$1.6 10$^{12}$ & $<$ 0.06\\
 \hline
                            &                                  &             & &     CSO           &         &\\
\hline
Position                      &  rms           &   $\delta$v              &  $\Delta$v    &   T$_{a}^*$  & &N(o--\hddp)$^\mathrm{b}_\mathrm{LTE}$ & tau$^\mathrm{b}_\mathrm{LTE}$\\
($^{\prime\prime}$)  & (K)            & km~s$^{-1}$  & km~s$^{-1}$ & K   &         &   cm$^{-2}$                 & \\  
\hline
(0,130)                    & 0.08           &  0.077       &               &$<$ 0.1&&  $<$ 3.0 10$^{12}$  & $<$ 0.1\\
(0,110)                      & 0.08           &  0.039       &   0.50      &   0.25  &&  8.0 10$^{12}$  & 0.28\\
(0,90)                      & 0.08           &  0.039       &   0.43      &   0.36  &&  1.0 10$^{13}$  & 0.42\\
(0,70)                      & 0.08           &  0.039       &   0.40      &   0.39  &&  1.1 10$^{13}$   &   0.47      \\
(0,50)                      & 0.10           &  0.039       &   0.36      &   0.64  &&  2.0 10$^{13}$  & 0.95\\
(0,30)                        & 0.08           &  0.039       &   0.50      &   0.48  &&  1.8 10$^{13}$   & 0.63\\
(0,0)                     & 0.10           &  0.039       &   0.41      &   0.66  & & 2.4 10$^{13}$     & 1.00\\
(0,-20)                     & 0.06           &  0.039       &   0.56      &   0.33  & & 1.2 10$^{13}$    &  0.38\\
(0,-40)                     & 0.10           &  0.039       &   0.48      &   0.30  &&  9.5 10$^{12}$    &  0.34\\
\hline
Position                     &  rms           &   $\delta$v         &  $\Delta$v    &   T$_{a}^*$ & N(p--\ddhp)$^\mathrm{c}_\mathrm{MC}$ & N(p--\ddhp)$^\mathrm{c}_\mathrm{LTE}$ & tau$^\mathrm{c}_\mathrm{LTE}$\\
($^{\prime\prime}$)  & (K)            & km~s$^{-1}$  & km~s$^{-1}$ & K                 &   cm$^{-2}$ &   cm$^{-2}$ & \\  
\hline
(0,0)                         &   0.07         &  0.042       &                     &  $<$  0.07 & $<$2.4 10$^{13}$  & $<$ 1.5 10$^{13}$      &  $<$ 0.48   \\
\hline
\end{tabular}
\begin{list}{}{}
\item[$^{\mathrm{a}}$] Column densities have been computed after averaging spectra at symmetrical distances from center. 
\item[$^{\mathrm{b}}$] Column densities and opacities have been computed with a beam coupling correction of 70\% for both JCMT and CSO.
\item[$^{\mathrm{c}}$] Upper limit column density and opacity have been computed with a beam coupling correction of 60\% at CSO.

\end{list}
\end{table*}

The 345 GHz (respectively 650 GHz) sidecab receiver with a 50 MHz acousto-optical spectrometer backend 
was used for all observations with an average velocity resolution of 0.04~\kmpsb 
(respectively 0.02~\kmps). At the observed frequencies 
of 372.421385(10) GHz for the \hddpb (1$_{10}$-1$_{11}$) and 691.660483(20) 
for the \ddhpb (1$_{10}$-1$_{01}$) lines \citep{amano05}, the CSO 10.4 meters 
antenna has a HPBW of about 20$^{\prime\prime}$ and 11$^{\prime\prime}$ respectively.
We performed a cut in declination across the main dust peak in \object{L183} between 
(0,-40$^{\prime\prime}$) and (0,130$^{\prime\prime}$) for \hddpb and concentrated on the reference position for \ddhp.
The system temperature was typically 1000 to 2000 K for \hddpb and 2500 to 3500 K for \ddhp.

The beam efficiency at 372 GHz (respectively 692 GHz) was measured on Saturn and Jupiter and 
found to be $\sim$ 60\% (respectively $\sim$ 40\%) for point sources. Pointing was monitored every 1.5 hrs and found to be 
better than 3$^{\prime\prime}$. If the emission is extended compared 
to the beam size of CSO, as appears to be the case for \hddpb then the efficiency is about 70\% at 372 GHz and 
60 \%  at 692 GHz. 
The data reduction was performed using the CLASS program of the GILDAS software developed at IRAM and the 
Observatoire de Grenoble and the LTE data analysis with CASSIS developed at CESR (http://www.cassis.cesr.fr). 

These CSO observations confirmed that the \hddpb emission was extended in this source as we could hypothesize from several tracers and we carried on the project at the JCMT, recently equipped with a new 16-pixel camera to be able to fully map the source in a reasonable amount of time.

\subsubsection{JCMT observations}

The JCMT observations were obtained during semester 07A in service mode, using the HARP-B 16 pixel camera (one pixel, located in a corner, was unavailable). A third of the observations was obtained in jiggle--chop\footnote{http://www.jach.hawaii.edu/software/jcmtot/het\_obsmodes.html}  mode and two thirds in position switch  (PSw) mode. The jiggle--chop mode appeared to be no faster, the displacement of the telescope in PSw mode seeming minor compared to other overheads, and because the jiggle--chop mode works in the Nyquist regime, each pixel receives much less integration time than in PSw mode. As adjacent pixels had the same off spectrum subtracted, the spatial average did not bring much improvement and we subsequently changed to position switch mode because deep integration on weak signal appears more important than Nyquist sampling for this work. In PSw mode, we made 2$\times$2 pointings to fill the gaps in the camera, thus achieving a full beam sampling. Two such sets were performed to cover the main dust peak and its northern extension (Fig \ref{intensmap}) with one pixel row overlap between the two.

Most of the observations were run in band 1 weather ($\tau_{225GHz} < 0.05$) while a few were done in band 2 weather leading to rapid degradation of the system temperature. The source was observed only above 40\degr\ elevation and the band 1 weather system temperature was in the range 500-1000 K depending on the pixels and on the elevation. To observe both \hddpb and \ndhpb \jqt (at  372.672509 GHz), we tuned the receiver half way between the two lines and used a frequency resolution of 61 kHz so that the backend could cover both lines at once. 

Data pre-reduction was done with the Starlink software (KAPPA, SMURF and STLCONVERT packages) and subsequently translated into CLASS format for final reduction.

\subsection{CO depletion and dust content}

All other observations used in this paper have been obtained and published previously. The dust content of L183 both in emission and in absorption has been described in \citet{Pagani03,Pagani04}. The source size is half a degree and contains a long filament extending $\sim$6\arcmin\ from north to south. Two peaks are clearly visible, one just south of the middle of the filament (which we call the main peak) with an opacity of $\sim$150 \Avb and a second one, 3\arcmin\ north of the first one (the north peak) with an estimated opacity of $\sim$70  \Av. These peaks have the characteristics of prestellar cores. Most of the filament have an opacity above 40 \Av.

Two large scale \cdhob and \cdsob maps obtained with the Kitt Peak 12-m telescope fail to trace the dense filament \citep{Pagani05}. It is now well established that this is due to depletion of CO onto grains. Surprisingly, the opacity at which the depletion begins ($\sim$20 \Av) is two times higher than what is usually observed in other clouds \citep[e.g.][]{Alves99,Kramer99,Bergin02} though it still appears at a density ($\sim$3 \pdix{4} \cc) that is a typical depletion density threshold \citep{Pagani05}. Possibly, the low density envelope where depletion has not yet occurred is very extended in this cloud (which is confirmed by its large influence on the \cdhob \jdu\ line intensity \citep{Pagani02}. The depletion factor in volume for CO has been estimated to be 43 on average \citep{Pagani05} on the line of sight of the main dust peak and is probably much higher in the inner part of this core where density is above 1~\pdix{5} \cc.

\subsection{\ndhpb and \nddp}

\ndhpb and \nddpb have been mapped at both low \citep[Kitt Peak 12-m,][]{Pagani05} and high (IRAM 30-m) resolutions. From the high resolution data, a strip crossing the main dust peak has been extracted and published \citep{Pagani07}. In that paper, we performed a detailed analysis of the \ndhpb and \nddpb emission with a Monte Carlo model treating exactly the hyperfine structure and line overlap of these species. We derived several physical properties, namely a maximum density of 2 \pdix{6} \cc, with a radial dependence proportional to r$^{-1}$ up to 4000 AU and proportional to r$^{-2}$ beyond, a kinetic temperature of 7 ($\pm$ 1) K, a slight depletion of \ndhpb in the inner 3000 AU of the core and a deuterium fractionation which is non-measurable at 10$^4$ AU ($<$ 0.03) and reaches $\sim$0.7 ($\pm$ 0.15) in the center. As far as we know this is the highest fractionation reported yet for a singly deuterated species. However this may not be exceptional when compared to the detection of triply deuterated species, like ND$_3$ \citep{Lis02,vdTak02} and CD$_3$OH \citep{Parise04} or to the fact that most reported fractionations are line-of-sight averages and are not derived from detailed profiles.

\section{Analysis}

\begin{table}
\caption[]{Source parameters\,: distance from the core center, H$_2$ density, \ndhpb abundance and \nddp/\ndhp ratio. The position is measured away from the PSC center along the R.A. axis \citep[from][]{Pagani07}.} 
\label{input}
\begin{tabular}{ccccl}
\hline
\multicolumn{2}{c}{Position}                      &  H$_2$ density           &   \ndhpb abundance   & \nddp/\ndhp \\
\cline{1-2}
(AU)  & (arcsec)                          &  (cm$^{-3}$)    &                           &   \\
\hline
  0		&0& 2.09 10$^6$	&	 2.40E-11	&	 0.69 $\pm$ 0.15\\
 1310	&12& 9.23 10$^5$  &	 8.50E-11 	&	0.42 $\pm$ 0.05\\
 2620	&24& 5.33 10$^5$	&	 1.10E-10 &	0.25 $\pm$ 0.02\\
 3930	&36& 3.22 10$^5$	&	 1.53E-10 &	0.16 $\pm$ 0.03\\
 5240	&48& 1.86 10$^5$	&	 1.27E-10 &	0.06 $\pm$ 0.02\\
 6550	&60& 7.08 10$^4$	&	 1.00E-10	&	 $\le$ 0.03\\
\hline
\end{tabular}
\end{table}

We present 3 models in this paper : a Monte Carlo model to compute \hddpb and \ddhpb line intensities, a chemical steady-state model and a chemical time-dependent model.

Compared to previous works, we benefit here from two new sets of coefficients and a large set of observations in a PSC. The \htpb + H$_2$ (+isotopologues) set of rate coefficients are extracted from the PhD work of E. Hugo in advance of publication and the \htpb (and deuterated counterparts) recombination rates have been computed for this work by V. Kokoouline and C. Greene and are presented in Appendix \ref{VK}. Rate coefficients, as computed by E. Hugo, describe all possible interactions between \htpb and H$_2$ isotopologues, including reactive and non-reactive, elastic and inelastic collisional rates, while recombination coefficients describe the dissociative recombination (DR) rates of trihydrogen cation isotopologues. Both sets take into account all the ortho, para and (\dtp) meta states.

\subsection{Line emission}

We have analysed the line emission of both ortho--\hddpb and para--\ddhpb using a two-level Monte Carlo code \citep[adapted from][]{Bernes79} with our new collisional coefficients. Because the temperature of the PSC is around 7 K for both the gas \citep{Pagani07} and the dust \citep{Pagani03,Pagani04}, the possibility of populating the next rotational level, \jkkb = 2$_{12}$ at 114 K and 75 K above ground level for ortho--\hddpb and para--\ddhpb respectively, is so low that it can be safely ignored. 
With the new coefficients, the critical densities are 1.1 $\times$ 10$^5$ and 4.9 $\times$ 10$^5$ \ccb respectively and thus the lines are close to thermal equilibrium in the inner core. In the case of ortho--\hddp, the difference between LTE ($\sim$2.0 $\pm$ 0.25 \pdixb{13}\sqc) and Monte Carlo ($\sim$2.3 $\pm$ 0.25 \pdixb{13}\sqc) column density estimates is typically 10--15\% in the direction of the dust peak. 
 
The para--\ddhpb line has not been detected \citep[see][]{Caselli08} and the 3 $\sigma$ upper limit corresponds to a total column density of 
$\sim$2.4 \pdixb{13}\sqcb using the Monte Carlo code. The LTE estimate yields $\sim$1.5 \pdixb{13}\sqc.

\begin{figure}[tbp]
\centering
\includegraphics[height=8.5cm,angle=-90]{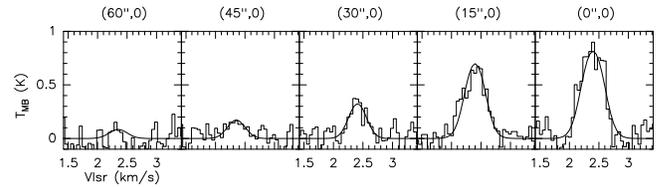}
\caption{ortho--\hddpb spectra across the main dust peak. East and west sides are folded together and fitted with a Monte Carlo model. Density and temperature profiles are taken from \citet{Pagani07}. The spacing between spectra is 15\arcsec}
\label{MCfit}
\end{figure}

\subsection{Deuteration}
%%\subsection{Description}

\htpb ions are formed at a rate 0.96\,$\zeta$ by cosmic ray ionization of H$_2$ \citep{Walmsley04}, rapidly followed by  reaction with another H$_2$ to form \htp,  and are destroyed in reactions with neutral species and in dissociative recombination with free electrons, negatively charged grains and possibly negatively charged polycyclic aromatic hydrocarbons (PAHs$^-$). In prestellar cores, the primary reservoirs of hydrogen and deuterium are 
H$_2$ and HD, with HD/H$_2$~=~2(D/H)$_{cosmic}$ $\sim$ 3.2 \pdix{-5} \citep{Linsky07}. 
The proton exchanging reaction of \htpb with H$_2$ partly regulates the H$_2$ ortho--to--para 
ratio but has no effect on the \htpb abundance. Concurrently, the reaction with HD forms \hddpb and this primary 
fractionation is then followed by the subsequent fractionations and produces \ddhpb and \dtpb 
\citep{Phillips03,Roberts03}\,:

\begin{eqnarray}
{\rm H_3^+ + HD} & \longleftrightarrow &{\rm H_2D^+ + H_2 + 232\,K}\\
{\rm H_2D^+ + HD}&  \longleftrightarrow &{\rm D_2H^+ + H_2 + 187\,K}\\
{\rm D_2H^+ + HD} & \longleftrightarrow& {\rm D_3^+ + H_2 + 234\,K.}
\end{eqnarray} 

\begin{figure}[tbp]
\centering
\includegraphics[height=6.7cm,angle=0]{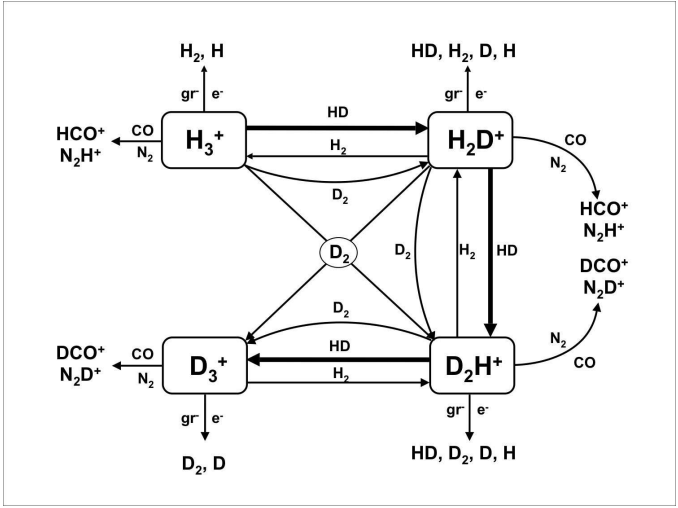}
\caption{Main reactions involved in the \htpb chemical network. When CO and N$_2$ are depleted, the reactions with bold 
arrows are dominant}
\label{fig:reactions}
\end{figure}

\noindent The backward reactions are endothermic with an energy barrier of about 200 K  (when considering only the ground level for each species) and 
were thought to be negligible at the low temperatures found in 
prestellar cores ($\leq$ 20 K), in which case the abundance ratios $\frac{[H_nD_{3-n}^+]}{[H_{n+1}D_{2-n}^+]}_{n=0,1,2}$ would be greatly enhanced.
However, such enhancement can be limited by various processes (see Fig. \ref{fig:reactions})\,:
\begin{itemize}
\item dissociative recombination of \htpb (and its deuterated counterparts) with free electrons or negatively charged grains (and PAHs$^-$ ?).
\item  reactions with "proton-friendly" molecules such as CO and N$_2$ which destroy the trihydrogen cations to produce HCO$^+$ and \ndhp.
\item ortho--H$_2$ which can react with the deuterated trihydrogen cation and remove the deuterium (see below). 
\end{itemize}

In this modeling, we introduce the backward reactions to equations (1), (2), and (3) as we distinguish between ortho, meta and para states of the 
different species. When these reactions are completely neglected, 
the deuteration fractionation is considerably enhanced and observations towards pre-stellar cores cannot be reproduced \citep{Roberts03}. Indeed, 
the importance of considering ortho and para states of various H/D carriers in the chemistry of the trihydrogen cation and isotopologues was first discussed by 
\citet{Pagani92} and subsequently expanded in a series of papers by Flower and coworkers 
\citep[][hereafter collectively referred to as FPdFW]{Walmsley04,Flower04,Flower06a,Flower06b}. 
\emph{Not only is this important in comparing the chemical model predictions on the abundance of \hddpb to the observations of the ortho--\hddpb 
species alone but also because some important reactions are much faster with ortho--H$_2$ than with para--H$_2$ hence no longer negligible}. Indeed, the 170K internal energy of the lowest ortho--H$_2$ level (J=1) is large with respect to the temperatures of interest and can significantly enhance the Boltzmann factor of endothermic reactions. In some cases, reactions which are endothermic with para--H$_2$ can turn out to be exothermic with ortho--H$_2$ i.e. fast and temperature independent. \emph{In fact, the ortho--to--para ratio of H$_2$ is found to be a crucial parameter for the entire chemistry of deuterium.}

The key reactions involving ortho--H$_2$ are essentially with meta--\dtp, para--\ddhpb and ortho--\hddpb as well as para--\hddp (see rates in Appendix A) because the internal energy of the ortho--H$_2$ alone is not enough to overcome the endothermicity of  reactions 1 to 3. Thus only those species which also have an internal energy high enough (so that the sum of the two internal energies is higher than the endothermicity of reactions 1 to 3) can react with ortho--H$_2$ at the Langevin rate in cold gas. Thus, the three reactions involving ortho--H$_2$ with  meta--\dtp, para--\ddhpb and ortho--\hddpb present exothermic or thermoneutral
channels to rehydrogenate the cations forming ortho--\ddhp, para--\hddpb and ortho/para--\htpb while the reaction between ortho--H$_2$ and  para--\hddpb can efficiently convert the latter to ortho--\hddp. The ortho--H$_2$ molecule thus opens a path to climb the 4 step energy ladder back from para--\ddhpb to \htpb via para--\hddpb and ortho--\hddpb which can be very efficient in the presence of large ortho--H$_2$ fractions. However, this efficient ladder scheme does not include \dtpb because the conversion of ortho--\ddhpb to para--\ddhpb is strictly forbidden in collisions with H$_2$ and very inefficient in collisons with HD. Nevertheless, these reactions can be a strong limit to the isotopic fractionation of \htpb hence of other species. Any chemical model which includes deuterium chemistry must distinguish between ortho and para states of dihydrogen and 
trihydrogen cation isotopologues and include reactions between the different spin states following \citet{Pagani92,Flower06a,Flower06b} and the present work.

\subsection{CO and N$_{2}$ chemistry}

The CO and N$_2$ abundances are critical parameters in the deuteration of the \htpb ion. CO is expected to freeze-out onto the grain 
mantles at high densities (a few 10$^4$ cm$^{-3}$) and low temperatures ($\leq$ 20 K) \citep[e.g.][]{Caselli99,Bergin01,Bacmann02,Tafalla02,Pagani05}. 
With an N$_2$ binding energy 
similar to the CO binding energy \citep{Oberg05,Bisschop06}, these two molecules are expected to behave similarly. However, observations towards 
prestellar cores prove that \ndhpb (produced from N$_2$) remains in the gas phase at higher densities than CO. 
This can be explained by the fact that \ndhpb is mainly
destroyed by CO \citep{Caselli02,Pagani05,Aikawa05}, so that the CO freeze-out implies a drop in the destruction 
rate of N$_2$H$^+$. This would partially balance the lower formation rate due to the N$_2$ freeze-out. Consequently, N$_2$H$^+$ is observed 
to survive in the gas phase at higher densities ($\sim$10$^6$ cm$^{-3}$). In the case of L183, we have shown that \ndhpb partially survives but suffers from  
depletion at densities starting at $\sim$5 \pdix{5} \ccb to reach a factor 6$^{+13}_{-3}$ at the core centre ($\sim$2 \pdix{6} \cc).  Because of growing deuterium 
fractionation, \nddpb abundance still increases towards the PSC center until the N$_2$ depletion becomes predominant over the deuterium enhancement, and in turn, the \nddpb 
abundance slightly decreases in the inner most part of the core \citep{Pagani07}.
 
The N$_{2}$D$^{+}$ and N$_{2}$H$^{+}$ ions can be produced via the following routes:
\begin{eqnarray}
%\begin{alignat}
\label{react0}
{\rm H_3^+ + N_2} &\rightarrow{\rm  N_2H^+ + H_2}\\
\label{react1}
{\rm H_2D^+ + N_2} &\rightarrow {\rm N_2D^+ + H_2} &\quad\quad\mbox{(for~1/3)}\\
\label{react2}
&\rightarrow {\rm N_2H^+ + HD} &\quad\quad\mbox{(for~2/3)}\\
\label{react3}
{\rm D_2H^+ + N_2}& \rightarrow {\rm N_2D^+ + HD}  &\quad\quad\mbox{(for~2/3)}\\
\label{react4}
%D_2H^+ + N_2 
&\rightarrow {\rm N_2H^+ + D_2}&\quad\quad\mbox{(for~1/3)}\\
\label{react5}
{\rm D_3^+ + N_2}& \rightarrow{\rm N_2D^+ + D_2.}
\end{eqnarray}
%\end{alignat}

\noindent We assumed that all the \htpb isotopologues react at the Langevin rate k$_{N_2}$ with N$_2$ (which is inversely proportional to the square root of the reduced mass of the colliding system, hence to the mass of the \htpb isotopologue) and that deuterium and hydrogen nuclei are equiprobably transferred. Consequently, \htp, \hddp, \ddhpb and \dtpb respectively produce an \nddp:\ndhpb ratio of 0:3, 1:2, 2:1 and 3:0. The measured ratio of 0.7 $\pm$ 0.15 in the center of L183 thus implies significant fractions of \ddhpb and \dtp. It has been shown \citep{Walmsley04} that in the case of complete depletion of heavy species (C, N, O...), \dtpb would be the dominant trihydrogen cation isotopologue which would imply that \nddpb is more abundant than \ndhp. This is not the case here\,; nevertheless the \ndhpb deuterium fractionation is a good constraint to the abundance of the four trihydrogen cation isotopologues in our chemical model. 

At steady state (d[N$_{2}$H$^{+}$]/dt=0 and d[N$_{2}$D$^{+}$]/dt=0), reaction \ref{react0} and its isotopic variants (\ref{react1} to \ref{react5}) 
being the main path to produce \nddpb and \ndhp, we obtain\,:
\begin{equation}
{\rm \frac{[N_2D^+]}{[N_2H^+]} = \frac{[H_2D^+]+2[D_2H^+]+3[D_3^+]}{3[H_3^+]+2[H_2D^+]+[D_2H^+].}}
\end{equation}
This ratio has been measured in the cut through the L183 main PSC. We describe in the following how our method can provide an estimate 
of the ortho/para \hddp ratio using this variable and subsequently, of the ortho/para H$_2$ ratio itself as well as some indication of the cosmic ionization rate 
and mean grain size.

\subsection{Grain distribution\label{grains_sec}}

Recombination of ions with electrons on negatively charged grain surfaces is an important process since it can be much faster than in the gas phase, especially in the case of H$^+$ 
\citep{Draine87}. 
The negatively charged grain surface area is therefore a crucial parameter \citep[we can safely ignore positively and multiply negatively charged grains, considered to be very rare in cold environments,][]{Draine87}. The grain size distribution in prestellar cores is unknown since it mostly 
depends on grain condensation and also on ice condensation \citep[e.g.][]{Vastel06}. 
We thus treat the grain radius a$_{gr}$ as a parameter of the model, assuming all the grains to have the same size and the dust to gas mass ratio to be 0.01. Different values could be advocated \citep[for example, in places where depletion is important, the ices increase the dust mass by up to 25\%,][]{Walmsley04} but the net result is to change only slightly the average grain size which is not well constrained in the PSCs in any case.

The focusing effect of the Coulomb attraction between charged particles and oppositely charged grains has been included 
using the \citet{Draine87} formalism:
\begin{equation}
\tilde{J}(Z=-1) = \left(1+\frac{1}{\tau}\right)\left(1+\sqrt{\frac{2}{2+\tau}}\right)
\end{equation}
where $\tau$ is the reduced temperature ($\tau=a_{gr}kT/e^2$, e being the electron charge, k the Boltzman constant). 
Therefore the recombination rate of the H$^+$ ion on a negatively charged grain can be expressed as:
\begin{equation}
k_{\rm gr}=\sqrt{\frac{8kT}{\pi m_{\rm H}}}\pi a_{\rm gr}^2(S \times \tilde{J}(Z=-1) ) 
\end{equation}
where a$_{\rm  gr}$ is the grain radius, $m_{\rm H}$ is the hydrogen mass, T the kinetic temperature and S is the sticking coefficient (S $\leq$ 1). The latter represents the probability that a 
colliding species will stick onto the grain surface. For ions, \citet{Draine87} concluded that the sticking coefficient should be unity. The same 
computation can be made to estimate the recombination rate of other ions, \htp, \hddp, \ddhp, \dtp,$\cdots$ by a simple correction of the atomic mass of the ions (respectively k$_{\rm gr}$/$\sqrt{3}$, k$_{\rm gr}$/$\sqrt{4}$, 
k$_{\rm gr}$/$\sqrt{5}$, k$_{\rm gr}$/$\sqrt{6}$, $\cdots$). 
In the case of collisions between charged particles and neutral grains, the attraction due to the polarization of the grain by the 
charged particle can be expressed through:
\begin{equation}
\tilde{J}(Z=0) = 1+\sqrt{\frac{\pi}{2\tau}}.
\end{equation}

\noindent Therefore the sticking rate of electrons on neutral grains can be expressed as:
\begin{equation}
k_{\rm e} = \sqrt{\frac{8kT}{\pi m_{\rm e}}}\pi a_{\rm gr}^2 (S \times \tilde{J}(Z=0))
\end{equation}
where $m_{e}$ is the electron mass and S is the sticking coefficient. S is about unity \citep{Umebayashi80} for a planar surface 
but a curvature of the grain surface will tend to reduce this parameter. However in the following we will use a factor of about unity as this parameter did not seem to have a large influence on the results in our runs.\\
The grain abundance [gr] can be expressed using:
\begin{equation}
[gr]  =   \frac{m_{\rm H_2}f_{\rm d/g}}{ \frac{4 \pi}{3} a_{\rm gr}^3 \delta}
\end{equation}
where $\delta$ is the mean grain density (assumed to be 3 g~\cc, $f_{\rm d/g}$ is the dust--to--gas mass ratio, and 
$m_{\rm H_2}$ is the mass of molecular hydrogen.
%\begin{equation}
%[gr] = 3.2~10^{-12}\left(\frac{f_{d/g}}{0.01}\right)\left(\frac{a_{gr}}{0.1 \mu m}\right)^{-3}
%\end{equation}
Another important parameter in our model is the abundance of the negatively charged grains ($[gr]=[gr^0]+[gr^-]$). At steady-state, assuming partial depletion of CO and N$_2$ and total depletion of all the other heavy species\,:
\begin{eqnarray*}
\rm{\frac{d[gr^-]}{dt}}& = &\rm [gr^0][e^-]{\it k}_e - [gr^-][H^+]{\it k}_ {gr} - [gr^-][H_3^+]{\it k}_ {gr3} \\
  & & \rm{} - [gr^-][H_2D^+]{\it k}_ {gr4} - [gr^-][D_2H^+]{\it k}_ {gr5}-[gr^-][D_3^+]{\it k}_ {gr6}\\
  &&\rm{}-[gr^-][N_2H^+]{\it k}_ {N_2H^+} -[gr^-][N_2D^+]{\it k}_ {N_2D^+}\\
  &&\rm{}-[gr^-][HCO^+]{\it k}_ {HCO^+} -[gr^-][DCO^+]{\it k}_ {DCO^+}  = 0
\end{eqnarray*}
(here we have neglected HD$^+$, D$_2^+$, He$^+$,...)
\subsection{Steady-state chemical model\label{steady}}

The code we describe in the following is used to calculate the steady-state abundances of the chemical species found in the different layers of the L183 prestellar core as listed in Table \ref{input}.

In the steady-state approximation the abundance species are interlinked via their production rates and their destruction rates 
(production=destruction).\\
Since H$_3^+$ is produced at a rate 0.96$\zeta$, the H$^+$ abundance can be expressed as (including only the main reactions)\,:\\
\begin{equation}
[\rm H^+] = \frac{0.04\zeta}{{\it k}_ {rec}[e^-]{\it n}_ {H_2}+{\it k}_ {gr}[gr^-]{\it n}_ {H_2}}
\end{equation}

The main production path of \htpb is via cosmic ray ionization of H$_2$ and proceeds in two steps :
\begin{equation}
\rm\zeta + o\textendash H_2  \rightarrow o\textendash H_2^+ + e^-
\end{equation}
\begin{equation}
\rm\zeta + p\textendash H_2  \rightarrow p\textendash H_2^+ + e^-
\end{equation}
and H$_2^+$ rapidly reacts with another H$_2$ to form H$_3^+$ but the branching ratios between different combinations of spin states are non-trivial \citep{Oka04} :

\begin{eqnarray}
\rm p\textendash H_2^+ + p\textendash H_2   \rightarrow p\textendash H_3^+ + H~~~~~~~~~~\\
\nonumber\\
\rm p\textendash H_2^+ + o\textendash H_2  \rightarrow  p\textendash H_3^+ + H~~~~~2/3\\
\rm 	   \rightarrow  o\textendash H_3^+ + H~~~~~1/3\nonumber\\
\nonumber\\
\rm o\textendash H_2^+ + p\textendash H_2  \rightarrow  p\textendash H_3^+ + H~~~~~2/3\\
\rm 	    \rightarrow  o\textendash H_3^+ + H~~~~~1/3\nonumber\\
\nonumber\\
\rm o\textendash H_2^+ + o\textendash H_2  \rightarrow  p\textendash H_3^+ + H~~~~~1/3\\
\rm 	    \rightarrow  o\textendash H_3^+ + H~~~~~2/3\nonumber
\end{eqnarray}
These are different from those advocated by FPdFW who took branching ratios of 1/2 for both species.
The ortho--H$_3^+$ formation rate from cosmic ray ionization k$_{cr\textendash o}$ is therefore the sum of several terms\,:
\begin{eqnarray}
k\rm_{cr\textendash o} = 0.96(1/3[p\textendash H_2][o\textendash H_2^+]+1/3[o\textendash H_2][p\textendash H_2^+]+\nonumber\\
 \rm        2/3[o\textendash H_2][o\textendash H_2^+])
\end{eqnarray}
The production rate for ortho--\htpb can be expressed as (including only the main reactions. The rates are listed in Table \ref{reactions})\,:
\begin{eqnarray}
k\rm_{cr\textendash o}\zeta + \it k\rm\_1_{oood}[o\textendash H_2][o\textendash H_2D^+]\it n\rm_{H_2}+(\it k\rm0_{poop}+\nonumber\\
         \it k\rm0_{pooo})[p\textendash H_3^+][o\textendash H_2]\it n\rm_{H_2}+\it k\rm1_{pdod}[HD][p\textendash H_3^+]\it n\rm_{H_2}
\end{eqnarray}
which represent respectively the formation from cosmic ray ionization, backward destruction of ortho--\hddp with ortho--H$_2$, spin conversion of para--\htpb with ortho--H$_2$ and finally, spin conversion of para--\htpb with HD.

The destruction rate for ortho--\htpb can be expressed as (including only the main reactions)\,:
\begin{eqnarray}
(o\textendash k\rm_{rec1}[e^-]+(\it k\rm0_{oopp}+\it k\rm0_{oopo})[o\textendash H_2]\nonumber\\+(\it k\rm1_{odpd}+{\it k}\rm 1_{odpo}+\nonumber
       {\it k}\rm 1_{odop}+{\it k}\rm 1_{odoo})[HD]\\+{\it k}\rm _{co}[CO]+{\it k}\rm _{N2}[N_2]+k_{gr1}[gr^-]))[o\textendash H_3^+]\it n\rm_{H_2} 
\end{eqnarray}
which respectively represents its destruction by dissociative recombination with electrons, spin conversion with ortho-H$_2$, spin conversion and deuteration with HD, proton transfer reactions with CO and N$_2$ and dissociative recombination on grains.

Similarly, the para--H$_3^+$ formation from cosmic ray ionization can be expressed as\,:
\begin{eqnarray}
k\rm_{cr\textendash p} = 0.96([p\textendash H_2][p\textendash H_2^+]+2/3[p\textendash H_2][o\textendash H_2^+]+\nonumber\\
\rm        2/3[o\textendash H_2][p\textendash H_2^+]+1/3[o\textendash H_2][o\textendash H_2^+])
\end{eqnarray}
The production rate for para--\htpb is\,:
\begin{eqnarray}
k\rm_{cr\textendash p}\zeta+({\it k}\rm \_1_{oopd}[\mathrm{o\textendash H_2}][o\textendash H_2D^+]+\nonumber\\
       ({\it k}\rm 0_{oopp}+{\it k}\rm 0_{oopo})[o\textendash H_3^+][o\textendash H_2]+{\it k}\rm 1_{odpd}[HD][o\textendash H_3^+]+\nonumber\\
       {\it k}\rm \_1_{popd}[o\textendash H_2][p\textendash H_2D^+])\it n\rm_{H_2}
\end{eqnarray}
and the destruction rate is\,:
\begin{eqnarray}
(p\textendash {\it k}\rm _{rec1}[e^-]+({\it k}\rm 0_{poop}+{\it k}\rm 0_{pooo})[o\textendash H_2]\nonumber\\+({\it k}\rm 1_{pdod}+{\it k}\rm 1_{pdpo}+\nonumber
       {\it k}\rm 1_{pdop}+{\it k}\rm 1_{pdoo})[HD]\\+{\it k}\rm _{co}[CO]+{\it k}\rm _{N2}[N_2]+{\it k}\rm _{gr1}[gr^-]))[p\textendash H_3^+]\it n\rm_{H_2} 
\end{eqnarray}

The N$_2$ abundance has been solved numerically to obtain the observed N$_2$H$^+$ abundance. 
Electronic abundance is adjusted to reach equilibrium.

In our steady-state model, the H$_2$ ortho/para ratio, the average grain radius and the cosmic ionization rate $\zeta$ are the varying input parameters. 
Within each layer of the PSC model (Table \ref{input}), these parameters are adjusted to match the following\,:
\begin{itemize}
\item the H$_2$ density
\item the \nddp/\ndhpb ratio at 7K
\item the observed ortho--\hddpb column density
\item the upper limit on the p--\ddhpb column density
\end{itemize}
Although the full range of grain sizes and ortho--to--para H$_2$ ratios have been explored for each H$_2$ density, we have not allowed solutions in which, for example, the grain size would oscillate from one layer to the next. We have searched for solutions throughout the layers in two different scenarios\,: (i) the grain size and the ortho-to-para ratio were both kept constant\,; or (ii) the grain size was allowed to increase and the ortho--to--para H$_2$ ratio to decrease with the H$_2$ density.  In both scenarios, $\zeta$ was kept constant throughout the 
layers. We neglected detailed reactions with D$_2$ as \citet{Flower04} have shown that its role is negligible in general and we have kept the HD 
abundance constant which is generally a good approximation.

\subsection{Time-dependent chemistry}

In a second step, we have constructed a pseudo time-dependent model based on NAHOON, a chemical model, a version of which has been made publicly available by 
V. Wakelam\footnote{http://www.obs.u-bordeaux1.fr/radio/VWakelam/Valentine\%20 Wakelam/Downloads.html}. We have modified 
this model in two ways : 1) we have replaced electron (resp. ion) reactions with neutral (resp. charged) grains as provided in the Ohio State University (OSU) reaction file 
(delivered with NAHOON) by the set of equations described above (Sect. \ref{grains_sec} and \ref{steady}), which we have directly included in the program, to take 
into account Coulomb focusing\,; 2)  we have included the formation of HD and D$_2$ on grain surfaces 
and we have introduced the spin state of H$_2$ and D$_2$ with the usual assumption that they are formed with the statistical ortho/para spin state 
ratio of 3 and 2 respectively. We have used the formation rate provided in the OSU reaction file (5\pdix{-17} cm$^3$\,s$^{-1}$) for the formation of molecular hydrogen. Because grains are covered by ice in the environments concerned here, we consider that the only
interaction between the atoms and the surface is physisorption. In this
case, the formation rates of HD and D$_2$ (in cm$^3$ s$^{-1}$) is half for
HD and 10$^5$ times lower for D$_2$ with respect to H$_2$ formation \citep[as dicussed in][]{
Lipshtat04}. In environments where grains are not covered by icy
mantles, on the other hand, one would have to consider chemisorption, which
strongly changes the efficiencies of the formation of HD and D$_2$ \citep{Cazaux08}
We have reduced the set of species and reactions to our needs, limiting ourselves to the most 
important reactions (see below) but differentiating all ortho and para (plus meta--\dtp) species as independent species and including all the detailed rates 
between the trihydrogen cation and dihydrogen isotopologues (see Sect. \ref{Hugo}) including spin state conversions. However we have 
included more reactions than in the steady-state model, taking into account reactions with D$_2$, H$_2^+$, He$^+$, etc. and allowing the ortho/para H$_2$ ratio and the 
HD abundance to vary. 

The main path to convert ortho--H$_2$ into para--H$_2$ is via the reaction
\begin{equation}
\rm o\textendash H_2 + H^+ \rightarrow p\textendash H_2 + H^+
\label{eq:o2pH2}
\end{equation}
which proceeds seven orders of magnitude faster at 7 K than the reverse reaction.

%H$_3$$^+$ + HD $\rightarrow$  H$_2$D$^+$ + H$_2$   &   k$_1$  & 1.3 10$^{-9}$  & 0 & 0 & Hugo \\
%o-H$_2$D$^+$ + o-H$_2$  $\rightarrow$ H$_3$$^+$ + HD  &  k$_{-1}$  &1.7 10$^{-9}$  & 0 & 0  & Hugo \\
%H$_2$D$^+$ + HD $\rightarrow$  D$_2$H$^+$ + H$_2$   & k$_2$   & 1.3 10$^{-9}$   & 0 & 0 & Hugo \\
%p--D$_2$H$^+$ + o-H$_2$ $\rightarrow$ H$_2$D$^+$ + HD   &  k$_{-2}$  & 1.7 10$^{-10}$  & 0 & 0  & Hugo \\
%D$_2$H$^+$ + HD $\rightarrow$  D$_3$$^+$ + H$_2$    &  k$_3$  & 1.3 10$^{-9}$  & 0 & 0  & Hugo \\
%D$_3$$^+$ + o-H$_2$ $\rightarrow$  D$_{2}$H$^+$ + HD   &  k$_{-3}$  & 1.4 10$^{-13}$  & 0 & 0  & Hugo\\

\subsection{Rate coefficients}

Many groups have made available gas-phase rate coefficients. The University of Manchester Institute of Science and Technology (UMIST) Database for Astrochemistry contains information on 4500 reactions of 
which 35\% have been measured experimentally, some at temperatures down to 20 K \citep{Woodall07}. The OSU group provides approximately 
the same database but focuses more on low temperature chemistry. We accordingly use in our modeling some of the reactions in the 
latter (with the most recent version OSU2007), considering the low temperatures found in L183. 
Apart from the rates presented below, all reaction rates involving deuterium have been taken from FPdFW (except recombination on grains for which we have used different sticking probabilities). We disregard the odd branching ratio of \ndhpb dissociative recombination reported by \citet{Geppert04} to consider a single possibility, namely the liberation of dinitrogen \citep{Molek07}.

\subsubsection{\htpb + H$_2$ isotopologue reaction rates\label{Hugo}}

Phase space theory (PST) was used to derive thermal state-to-state rate coefficients for the whole H$_3^+$ + H$_2 \rightarrow$ H$_3^+$ + H$_2$ system and isotopic variants in the temperature range 5--50K. This statistical method accounts for such quantities as mass, energy, rotational angular momentum, nuclear spin symmetry and their respective conservation laws. The ergodic hypothesis which is a requisite for PST as well as the full-scrambling hypothesis are assumed according to the topology of the PES \citep{Yamaguchi87,Xie05}. Reactant (product) trajectories are treated with the classical Langevin model. The resulting set of state-to-state rate coefficients deviates from the detailed balance principle by a few percent at worst and is consistent with thermodynamical equilibrium constants. Details will be given in a forthcoming publication (Hugo et al., in prep.)

In the present astrochemical model, nuclear spin states of the different molecules are treated as distinct species but their rotational states are not considered individually. We thus made the assumption that only the rotational ground states of each nuclear spin species were populated and used the ground state-to-species thermal rate coefficients obtained by summing ground state-to-state thermal rate coefficients over the product channels.

\subsubsection{\htpb isotopologue dissociative recombination rates}

The dissociative recombination (DR) rate of \htpb is difficult to measure experimentally. Since the early measurements e.g. 
by \citet{Adams84}, numerous attempts have been made and are summarized in two papers \citep{Bates93,santos07}. In parallel, theoretical work has also been 
developed with the latest one published by \citet{santos07}. Extending upon that work, we present here, in Table \ref{DR},  the \htpb updated DR rate 
\citep{santos07} along with newly calculated \hddp, \ddhp, and \dtpb DR rates (see appendix B for more details). These calculations do not predict the branching 
ratio of the DR products. We have thus adopted the branching ratios published elsewhere \citep{Datz95a,Datz95b,Strasser04,Zhaunerchyk08} which we have 
applied to the calculated rates. The resulting DR rates at 7 K are listed in Table \ref{reactions}.
Several remarkable effects at low temperature are visible (see Figs \ref{fig:thermal-rate_1}, \ref{fig:thermal-rate_2}, \ref{fig:thermal-rate_3}, \ref{fig:thermal-rate_4}) :

\begin{itemize}
\item the strong departure of the ortho--\htpb DR rate from that of para--\htp. The difference is a factor of 10 at 10 K.
\item \ddhpb shows a large DR rate drop, by a factor of 10 at 10 K for both ortho and para species compared to the extrapolated value used by FPdFW.
\item On the contrary, a large increase of the \dtpb DR rate is predicted to occur but mostly at temperatures where deuteration is low and therefore the consequences for \dtpb abundance are limited.
\end{itemize}

\section{Results and discussion}

\subsection{\hddpb linewidth}

In order to fit the observed ortho--\hddpb line profiles, we run the Monte Carlo model using the "best model" velocity, density and temperature profiles derived 
from the \ndhpb and \nddpb data analyzed in \citet{Pagani07}. However, the linewidth for the two stronger spectra (offsets (0,0) and (15\arcsec,0), Fig. \ref{MCfit}) 
is too wide to be reproduced with the same micro--turbulent width which we have used for \ndhp ($\Delta$v$_\mathrm{turb}$(FWHM) $\approx$ 0.14\,km\,s$^{-1}$). 
Indeed, the central spectrum linewidth measured by fitting a Gaussian yields 0.45 ($\pm0.03$) \kmps. The thermal contribution is
\begin{equation}
\Delta \mathrm{v_{therm}(FWHM)} = 2.336 \times \sqrt{\frac{kT}{m}} = 0.28~\mathrm{km~s^{-1}}
\end{equation}
at 7 K, which implies that $\Delta$v$_\mathrm{turb}$(FWHM) contribution is 0.35 \kmps, 2.5 times larger than for \ndhp. Here $k$ is the Boltzmann constant and $m$ is the mass of \hddp. If we impose a turbulent velocity similar to the one modeled for \ndhp, then the temperature needed to obtain such a wide line would be 16K which is completely ruled out by \ndhpb observations \citep{Pagani07}. Infall motion limited to the inner core could have explained the \hddpb width if no \hddpb spectrum other than the central one had been wide combined with a large depletion of \ndhp. This is not possible because the \hddpb spectrum at (15\arcsec,0) has the largest width (0.49 $\pm0.03$ \kmps) in a region where \ndhpb is hardly depleted. This remains therefore a pending problem.

\subsection{\ndhpb deuteration}

\begin{figure*}[tbp]
\centering
\includegraphics[height=17cm,angle=-90]{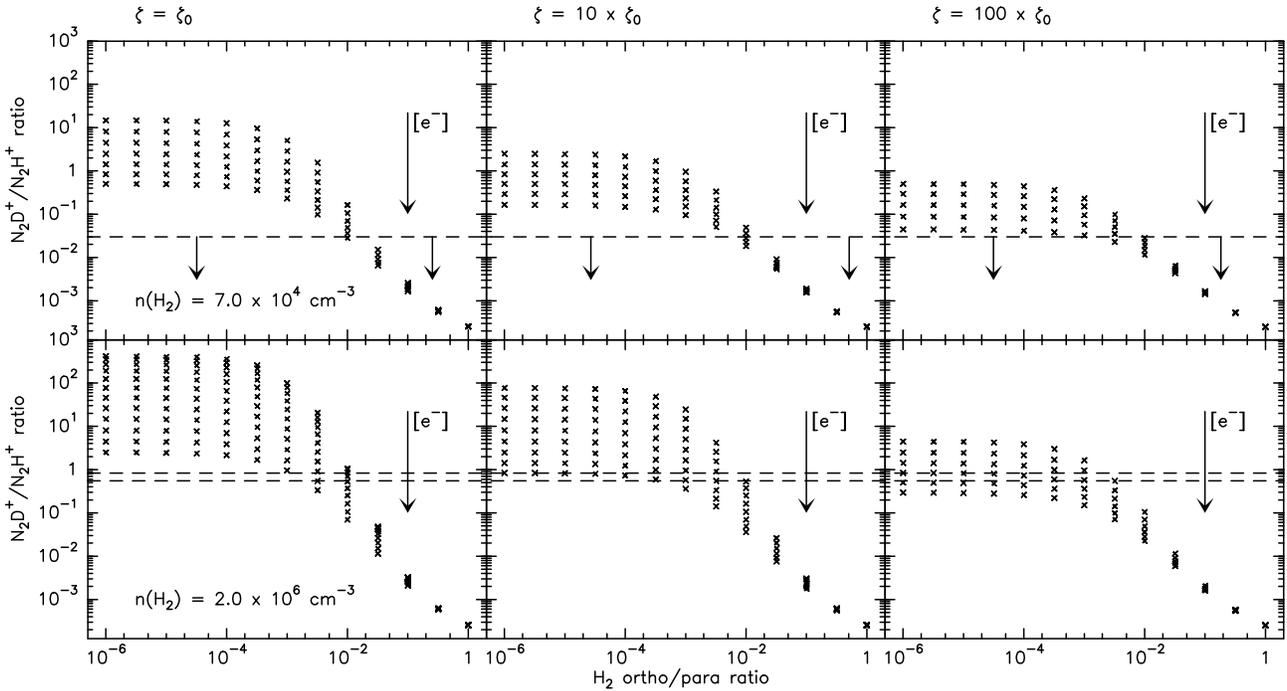}
\caption{\nddp/\ndhpb ratio as a function of the ortho/para H$_2$ ratio for all possible electronic abundances and total depletion (the steady-state chemical model). The lower row corresponds to the densest part of the PSC (n(H$_2$) = 2 \pdix{6} \cc) and the two horizontal dashed lines the measured range of the \nddp/\ndhpb ratio while the upper row corresponds to the external part of the PSC (n(H$_2$) = 7 \pdix{4} \cc) with the dashed line representing the  \nddp/\ndhpb ratio upper limit. The three columns represent different $\zeta$ values as indicated above ($\zeta_0$ = 1 \pdix{-17} s$^{-1}$) . The large arrow indicates the direction of increasing electronic abundance}
\label{n2hpratio}
\end{figure*}

\begin{figure*}[tbp]
\centering
\includegraphics[height=17cm,angle=-90]{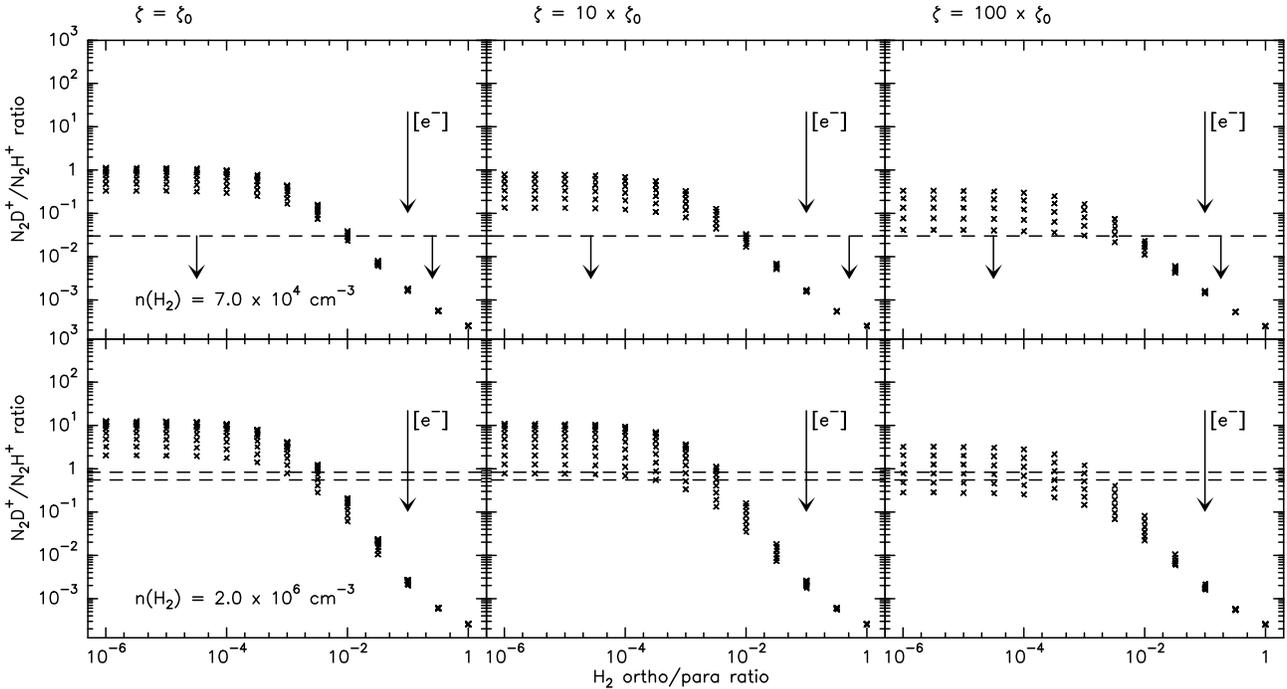}
\caption{Same as Fig. \ref{n2hpratio} but with a CO/H$_2$ abundance of 10$^{-5}$ in the outer layer (n(H$_2$) = 7 \pdix{4} \cc) and 10$^{-6}$ in the inner layer (n(H$_2$) = 2 \pdix{6}Ê\cc)}
\label{n2hpratioCO}
\end{figure*}

\subsubsection{Requested conditions}

We next discuss the main parameters that control the \ndhpb deuteration using the steady-state chemical model.

The models have been run for a temperature of 7 K which prevails in all the layers where \nddpb has been detected in the PSC cut presented in \citet{Pagani07}. We have also run the models for the corresponding density, \ndhpb abundance and \nddp/\ndhpb ratio of each layer (the parameters are listed in Table \ref{input}).

As discussed above, the abundance of ortho--H$_2$ is the main controlling factor of the trihydrogen cation isotopologue abundances and therefore of the \nddp/\ndhpb ratio \citep[and similarly of the DCO$^+$/HCO$^+$ ratio, see e.g.][]{Maret07}. We have therefore explored the range of possible solutions for the ortho/para H$_2$ ratio in the two extreme layers of our core profile (n(H$_2$) = 7 \pdix{4} \ccb and 2 \pdix{6}Ê\cc) for which we have a \nddp/\ndhpb ratio of $<$0.03 and 0.7 $\pm$ 0.15 respectively (Fig. \ref{n2hpratio}). We have done this for three cosmic ray ionization rates (10$^{-17}$, 10$^{-16}$, and 10$^{-15}$ s$^{-1}$) covering the values generally discussed in the literature \citep[e.g.][]{Maret07} and for all possible electronic abundances (or average grain sizes as they are linked via the abundance of H$^+$ which is mostly controlled by the grain surface area). In this first run, we have simulated total depletion by adjusting the CO and N$_2$ abundance\footnote{ Here, we look for general solutions of the  \nddp/\ndhpb ratio as a function of several parameters, we thus do not try to fit the N$_2$ abundance to obtain the observed \ndhpb abundance} to 10$^{-8}$. We have also indicated the range of \nddp/\ndhpb ratio measured in both layers. The average grain radius has been varied from 0.01 $\mu$m to 5 $\mu$m and electronic abundance from 10$^{-11}$ to 10$^{-6}$ which cover the usually accepted values. We can see that \nddp/\ndhpb ratios above 100 are possible in dense gas though they require very low electronic abundances and therefore very small grains which are probably absent from these dense and cold regions due to grain coagulation \citep[see e.g.][]{Stepnik1}.
 
In the lower density outer layer where no \nddpb has been detected, the ortho/para H$_2$ ratio must be high enough, i.e. above $\sim$0.01,  to prevent any deuteration from occuring whatever the cosmic ray ionization rate. On the contrary, the dense, strongly deuterated layer has solutions only below a maximum ortho/para H$_2$ ratio of 0.01 (or lower for high $\zeta$ rates). Thus the ortho/para H$_2$ ratio across the PSC clearly must vary from above 0.01 to below 0.01.
In the case of low cosmic ray ionization rates (10$^{-17}$ s$^{-1}$), though the ortho/para H$_2$ ratio of 0.01 seems to be a common solution for both layers, it requires a high electronic abundance (and large grains) in the outer layer and a low electronic abundance (and small grains) in the inner dense part. This is clearly improbable. 
The temperature being low enough in all the layers, warm layers (above 20 K) cannot be invoked instead of a high ortho/para H$_2$ ratio to limit the deuteration in the outer parts of the PSC. CO total depletion is however questionable and  in the model we also used a CO depletion factor of 10 (abundance of 10$^{-5}$) in the outer layer and a CO depletion factor of 100 (abundance of  10$^{-6}$) in the inner layer (Fig. \ref{n2hpratioCO}). This only limits the maximum \nddp/\ndhpb ratio which decreases by one order of magnitude. Indeed, the destruction of \htpb by CO dominates over recombination with electrons when their abundance is very low and vice versa. However, the conditions to reach the observed \nddp/\ndhpb ratio remain unchanged and therefore \emph{only a variable ortho/para H$_2$ ratio can be invoked}. Such a variable ortho/para H$_2$ ratio cannot be investigated with a steady-state model because in all layers, the ortho--H$_2$ abundance would eventually decrease to values about of 10$^{-3}$--10$^{-4}$ as discussed by FPdFW. %le 2006
%As far as we know, this is the first time that a variation of the ortho/para H$_2$ ratio is evidenced in a cloud.

\subsubsection{The ortho/para H$_2$ variation}

We discuss here the possibilities of making the ortho/para H$_2$ ratio vary across a single PSC.

It is commonly accepted that H$_2$ is formed on grain surfaces with an ortho/para ratio of 3 because of spin statistics and the exothermicity of the reaction H + H $\rightarrow$ H$_2$. Subsequently, the ortho--H$_2$ is converted into para--H$_2$ following equation \ref{eq:o2pH2} and to a lesser extent with reactions involving \htpb and its isotopologues. As already discussed by FPdFW, this conversion is slow and has probably not reached steady state in clouds with ages between 10$^{5}$ and 10$^{6}$ years. 

\begin{figure*}[tbp]
\centering
\includegraphics[height=17cm,angle=-90]{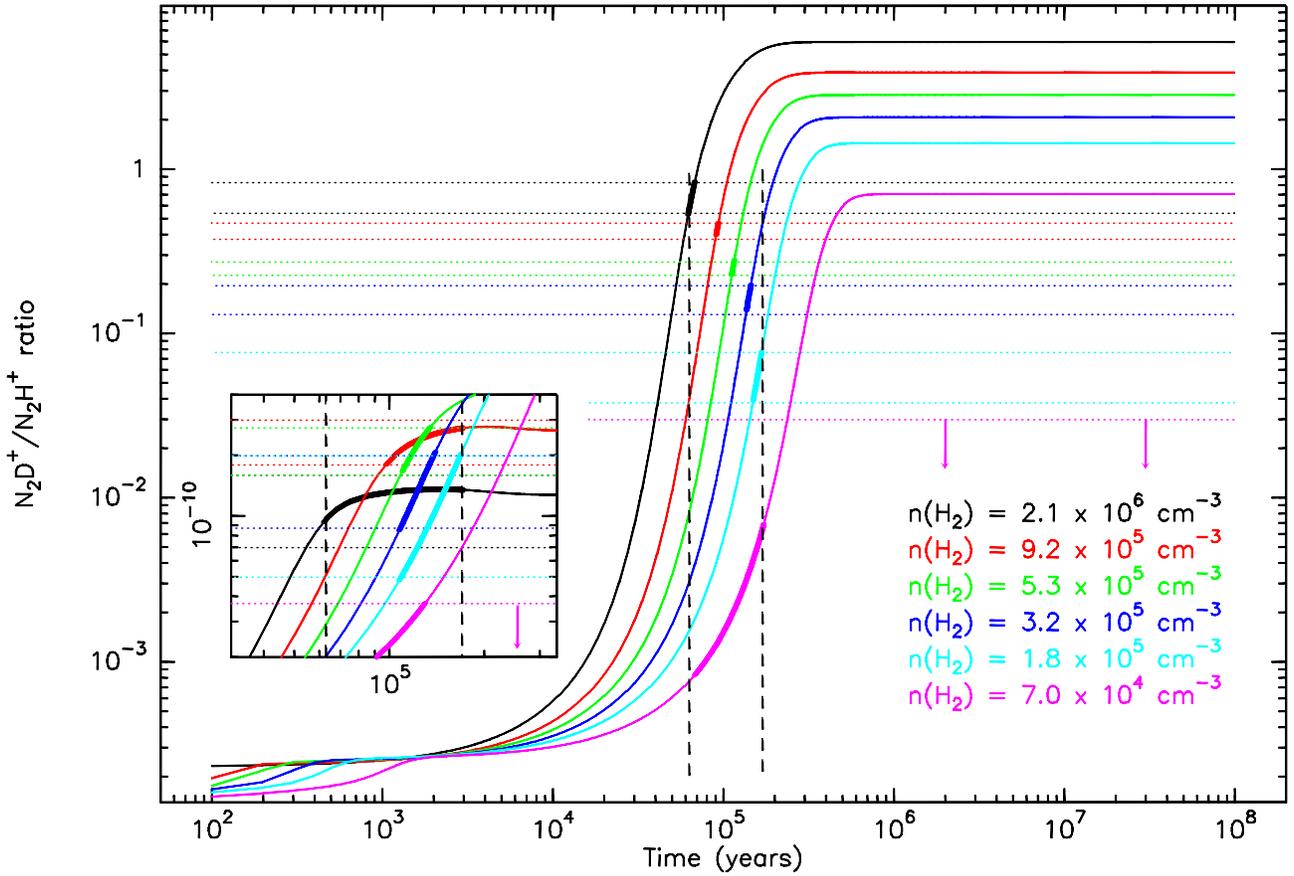}
\caption{Time-dependent variation of the \nddp/\ndhpb ratio for the 6 layers defined in Table \ref{input}. For each colour, the density of the layer is given. In the insert, the ortho--\hddpb abundance is represented with the same color code, zoomed on the epoch of interest. Horizontal dotted lines represent the observed \nddp/\ndhpb ratio range for each layer and the observed ortho--\hddp abundance as derived from the Monte Carlo model applied to the JCMT observations. The part of the chemical solution that fits in both these limits and the common time limits is set in bold. Vertical arrows indicate upper limits for the \nddp/\ndhpb ratio. Vertical dashed lines are placed at 0.63 and 1.7 \pdix{5} years to delimit the period when all layers reach their observed \nddpb enrichment. This case has been computed for an average grain radius of 0.1 $\mu$m and $\zeta$ = 2 \pdix{-17} s$^{-1}$}
\label{nahoon}
\end{figure*}

\begin{figure*}[tbp]
\centering
\includegraphics[height=17cm,angle=-90]{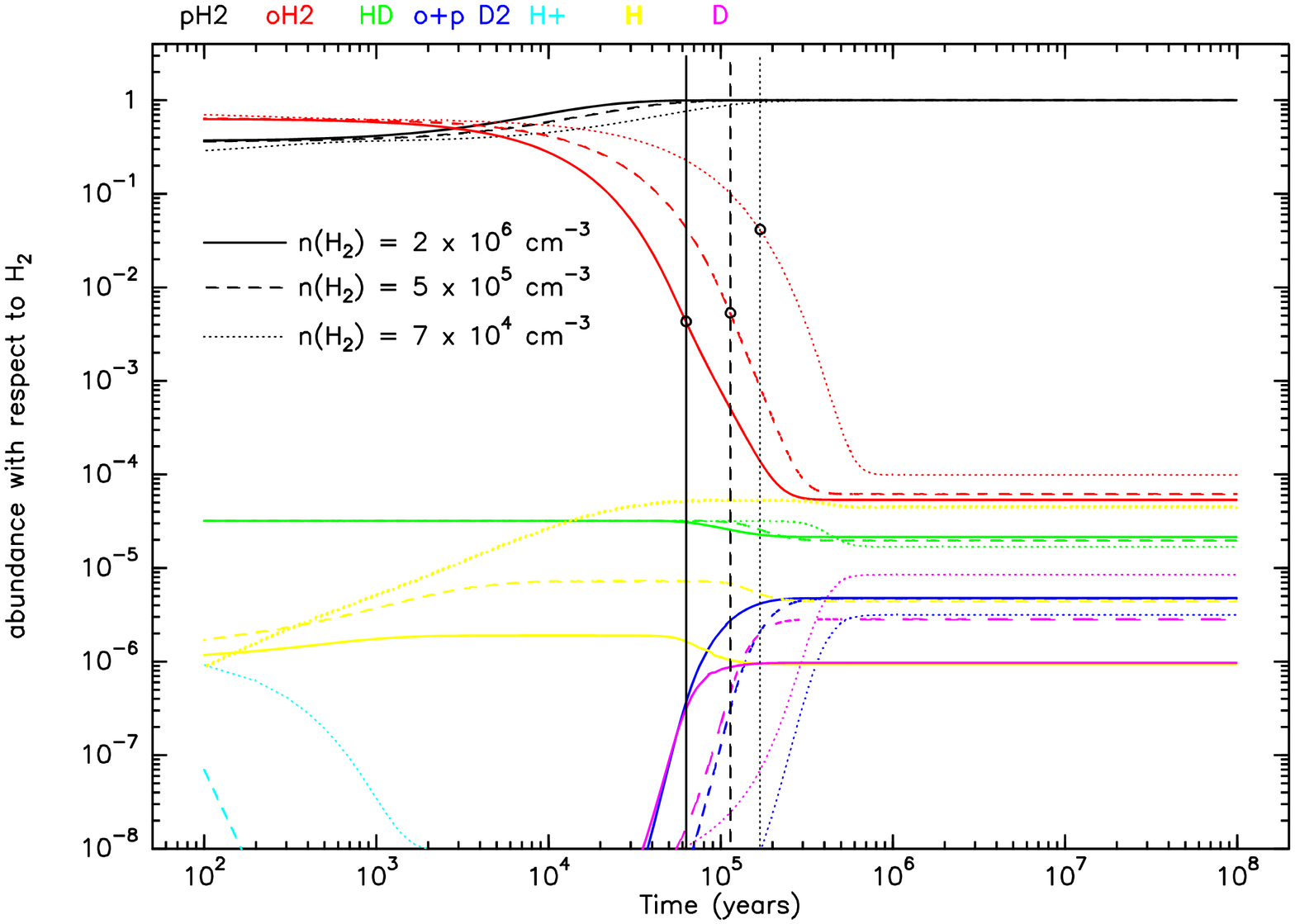}
\caption{Time-dependent variation of the ortho-- and  para--H$_2$ and other  related species for 3 of the 6 layers defined in Table \ref{input} (layers 1, 3 and 6) . Vertical lines are placed at the times corresponding to the observed \nddp/\ndhpb ratio in Fig. \ref{nahoon} for the same 3 densities. This case has been computed for an average grain radius of 0.1 $\mu$m and $\zeta$ = 2 \pdix{-17} s$^{-1}$ }
\label{nahoonH2}
\end{figure*}

We have therefore run the modified Nahoon model to search for a simultaneous solution for all layers which would get the necessary ortho/para H$_2$ gradient across the PSC to achieve both the observed \nddp/\ndhpb ratio and ortho--\hddpb abundance profiles. In this model, we have no direct measurement of the CO abundance but for the credibility of the model, we have set the CO abundance to 10$^{-5}$ in the outer layer  and increased the depletion to reach a factor of 100 (i.e. a CO abundance of 10$^{-6}$) in the core. The results are presented in Figs. \ref{nahoon} \& \ref{nahoonH2}. We have searched for a solution where each modeled layer meets the two observational constraints (\nddp/\ndhpb ratio and ortho--\hddpb abundance) at the same time, but these solutions are not simultaneous between the different layers. We varied the average grain radius
and the cosmic ray ionization rate, $\zeta$. We could find solutions for grains of average radius 0.025 to 0.3 $\mu$m . No solution has been found for grains above 0.3 $\mu$m.
Figs. \ref{nahoon} \& \ref{nahoonH2} show the case for which the grain average radius is 0.1 $\mu$m and $\zeta$ = 2 \pdix{-17} s$^{-1}$. In this case, the time range inside which all layers meet the requested conditions is 0.6 to 1.7 \pdix{5} years.  Figure \ref{nahoonH2} shows how the ortho/para H$_2$ ratio evolves for 3 selected layers. We have marked the appropriate time which is the solution for each of these layers as established from Fig. \ref{nahoon}. In that figure, we can see that the o/p H$_2$ ratio is below 0.01 for the dense layers and still above 0.01 for the outer layer for which no \hddpb has been detected, as expected from the steady-state model. We can also see that the full o/p H$_2$ relaxation has not yet occurred even for the densest part of the cloud. Smaller grains have a larger interacting surface and therefore lower the abundance of H$^+$ ions which preferentially recombine on negatively charged grains (or PAHs$^-$), consequently slowing down the dominant ortho--H$_2$ relaxation reaction (eq. \ref{eq:o2pH2}). Though smaller grains also imply a lower electronic density, therefore favouring a higher deuteration of \ndhpb as shown in Fig. \ref{n2hpratio}, the slower disappearance of ortho--H$_2$ is the dominant process here and finally, smaller grains slow down the deuteration process. For grains of average radius 0.025 $\mu$m, the range of ages matching the range of \ndhp/\nddpb observed ratios is 2.7--3.8 \pdixb{5} years, while for grains of average radius 0.3 $\mu$m, the time range is only 3.8--7.2 \pdixb{4} years. Figure \ref{nahoonH2} suggests that D$_2$ should become a sizeable fraction of available deuterium a short while after the present state (typically 2--3 \pdixb{5} years) and that HD correspondingly  should drop slightly.

\subsection{Age of the core and collapse}

Though it is normal that dense layers evolve faster than less dense ones, at least to account for a differential ortho/para H$_2$ ratio, the layers are evolving too fast in our model. The densest layers would have reached their present status 2 to 3  times faster than the outer, less dense layers. This could be possible only if the denser layers had reached their steady-state equilibrium. This is not the case here where the densest layer would reach its steady-state equilibrium only after 2 \pdix{5} years and this would imply a \nddp/\ndhpb ratio of 6, almost an order of magnitude higher than observed. The most probable reason for this time discrepancy is that the core has undergone a contraction and therefore all layers were not so dense in the past. While the outer layers have only little evolved in density (the most external one probably started at 0.5--1 \pdix{4} \ccb to reach 7 \pdix{4} \ccb today), the inner ones have undergone a much greater density increase. As constant density through the core would give no chemical differentiation while a time-frozen density profile as measured here gives too much differentiation, the solution is in between the two. Starting from a uniform gas, the chemical differential evolution of the core should therefore help us to constrain the duration of the contraction and the type of contraction. Of course the model should also include the evolution of depletion which also plays a role in the  acceleration of the process.

As the core must have started to contract from a lower density region, typically 10$^4$ \cc, it is clear that all layers have accelerated their chemical evolution while their density was increasing. Therefore, the layer with the longest time to reach the observed \nddp/\ndhpb ratio gives a lower limit to the age of the cloud. Depending on the exact average size of the grains, this lower limit is 1.5--2 \pdix{5} years here. It is even larger because before the cloud underwent contraction, depletion had not yet occured and therefore species like atomic sulfur, S, must have been present in quantities large enough to transfer notable quantities of electric charges from H$^+$ to S$^+$ (H$^+$ + S $\to$ H +  S$^+$)  and PAHs must have also been abundant enough to help destroy H$^+$ ions \citep{Wakelam08}.
All these phenomena contribute to the diminution of the H$^+$abundance, therefore slowing down the ortho--H$_2$ relaxation process. Indeed, \citet{Flower06b} show that the relaxation process in some cases takes 3 \pdixb{7} years, typically 50 times slower than in the case presented here (and 15 times slower for similar conditions of grain size and cosmic ray ionization rate but without depletion in their case).

\subsection{Para--\ddhp}
At 7 K, the strongest possible line intensity (LTE case) for the ground transition of para--\ddhpb is below 0.3 K because of the Rayleigh-Jeans correction at 691 GHz which becomes large. Moreover the thermalization of the line is difficult to obtain in this source beyond the radius of 3000 A.U. because the PSC density drops below the para--\ddhp critical density (n$_{crit}$ = 4.9 \pdix{5} \cc) and a slight drop of the excitation temperature turns into an exponential decrease of the brightness temperature.  For T$_{ex}$ = 6 K, T$_{bright} \leq$ 0.13 K. Therefore searches for para--\ddhpb must reach very low noise levels to have a chance of detection. 
From the models we present here, we predict an integrated line intensity of 11 mK km/s (34 mK peak) with the Monte Carlo model, and an upper limit of 16 mK km/s (48 mK peak) in the case of LTE. This is a factor of 3 to 4 below the upper limit we obtained from the observations.

\subsection{The chemical profile}

\begin{figure*}[tbp]
\centering
\includegraphics[height=24cm]{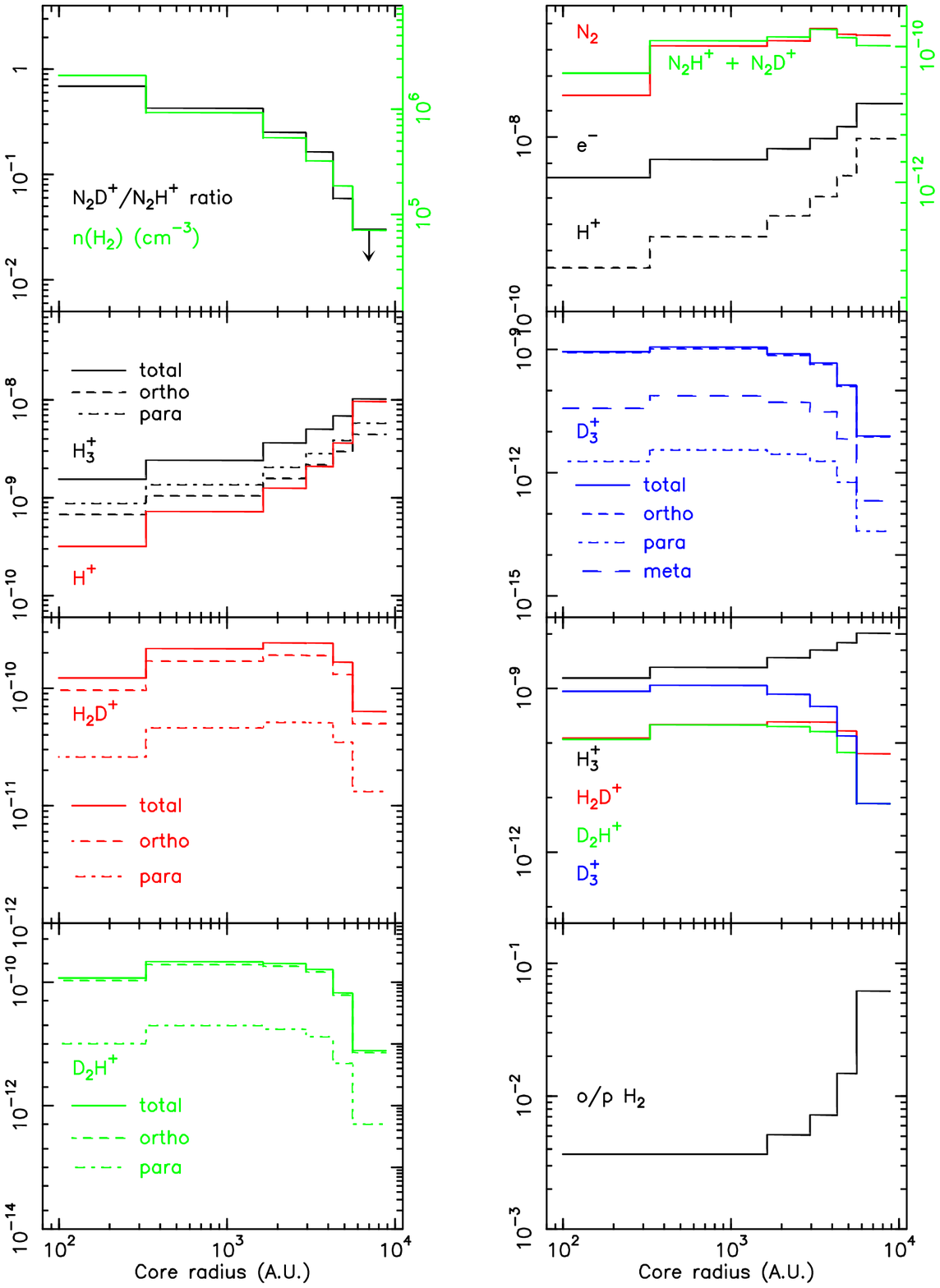}
\caption{PSC profile for the different species. The n(H$_2$), \ndhpb abundance, and \nddp/\ndhpb ratio profiles are input data. The four trihydrogen cation isotopologue profiles are also grouped together to visualize their relative total abundances and H$^+$ is compared to both e$^-$ (upper right box) and to \htpb (2nd upper left box). The profile has been computed for the case presented in Fig. \ref{nahoon}, i.e. $\zeta$ = 2 \pdix{-17} s$^{-1}$, a$_{gr}$ = 0.1 $\mu$m. In the two upper boxes, the green curve refers to the green axis on the right. In the top left box, the arrow indicates the \nddp/\ndhpb ratio upper limit for that layer}
\label{profile}
\end{figure*}

Finally, we obtain a detailed profile of the PSC which we present in Fig. \ref{profile}. It represents the solution  for the model we presented in Figs. \ref{nahoon} \& \ref{nahoonH2} taking for each layer the values at their respective best-fit time.
The large variation of the ortho--H$_2$ species across the core (a factor of 15) makes \dtpb change by an even larger amount (2 orders of magnitude) but it does not become the most abundant trihydrogen cation isotopologue in the core center at this stage because the ortho-H$_2$ abundance is not yet low enough. In the present case, for a density of 2 \pdix{6} \cc, the inversion between \htpb and \dtpb occurs when the abundance of ortho-H$_2$ drops below 3 \pdix{-3}.

The ortho/para H$_2$ ratio in the outer layer is 0.04 (as expected from Fig. \ref{n2hpratioCO} which indicates a lower limit of 0.01). As discussed above, the ortho/para H$_2$ ratio evolution speed is linked to density and grain size, both of which are lower outside the PSC, in its embedding parental cloud. We can thus expect this ratio to be at least 0.05 and probably above 0.1 in the envelope of the cloud. 

N$_2$ is an input parameter in our model because we have not included all the nitrogen chemistry. As reactions with \htpb isotopologues are the main path to destroy this molecule \citep{Flower06b}, we do not introduce a large error in determining its abundance directly from the \ndhpb abundance itself and obtain a N$_2$ profile which is probably closer to reality than if we had let the whole N-chemistry freely establish its abundance because of too many unknowns. The abundance profile thus starts at 1.5 \pdix{-7} with respect to H$_2$ in the low density layer to diminish to 3 \pdix{-8} in the densest layer. The undepleted N$_2$ abundance after attaining the steady state is $\sim$3 \pdix{-5} \citep{Flower06b} but this is reached only after $\approx$ 5 \pdix{6} years. As depletion of N$_2$ in the outer layer is possibly still small, we can conclude that N$_2$ has not yet reached its steady state abundance which puts an upper limit on the age of the cloud of about 1 \pdix{5} years following the estimate of \citet{Flower06b}. However, this depends very much upon several factors, for example the C:O elemental abundance ratio. Consequently this information is only indicative. 

Though our model does not deal with the nitrogen chemistry, it seems to indicate that low abundances of N$_2$ are sufficient to explain the observed \ndhpb abundance and therefore, the N$_2$ depletion seems not to be a critical factor as long as CO is also depleted. The much debated contradiction between the presence of \ndhpb in depleted cores while N$_2$ should deplete like CO would thus be a false problem. Low CO and electronic abundances, limiting the destruction rate of \ndhp, seem to be sufficient to compensate for the N$_2$ depletion itself to a large extent. Finally, in the inner core, N$_2$ and \ndhpb (+ \nddp) follow a similar decreasing trend, suggesting that N$_2$ depletion eventually forces \ndhpb to decrease. 

While ortho--\hddpb is 83 K above the para ground state, it is more abundant all over the core profile by almost an order of magnitude whereas the thermal equilibrium ratio would be ortho/para~\hddpb $\approx$ 2 \pdix{-5} at 7 K. This demonstrates the efficiency of the ortho--H$_2$ 
in converting para--\hddpb into ortho--\hddp and therefore limits the total abundance of \hddp, as the backward channel to \htpb remains open even at 7 K.
This ortho/para population inversion does not occur for \ddhpb as the species needed to perform this inversion is no longer ortho--H$_2$ but the much rarer ortho--D$_2$. Therefore, the para--\ddhpb remains the least abundant of the two spin state species which, combined with the fact that its ground transition is higher in frequency than the one of ortho--\hddp, makes its detection extremely difficult.

\section{Conclusions}

We have presented a pair of simple chemical models restricted to H-carriers, He plus CO and N$_2$ to account for the observed HCO$^+$, DCO$^+$ 
(not discussed in this paper), \ndhp, and \nddpb ions. We have benefited from new computed reactions rates for both the \htpb + H$_2$ isotopologue 
combinations and for the \htpb isotopologue dissociative recombination rates which explicitly take into account the nuclear states individually. 

With the steady-state model we have shown that the ortho/para ratio of H$_2$ must vary from above 0.01 in the outer parts of the L183 PSC to less 
than 0.01 in the inner parts to explain the variation of deuteration across the core.  Checking the reality of 
the ortho/para H$_2$ variation with a time-dependent model , we have also found that if the present PSC density profile is static, then the inner layer would have reached its present status 2 to 3 times faster than the outer layers. Because the present status is not in steady-state, the layers should evolve at a similar rate and therefore the density must have been lower in the past. The most probable explanation is that the core has probably evolved from a uniform density cloud to the present centrally condensed PSC. The time-dependent model also suggests that the ortho/para H$_2$ ratio changes by one order of magnitude from $\sim$5\% at a density of 7 \pdix{4} \ccb down to a few \pdix{-3} in the inner dense core. This has two important consequences\, :

\begin{itemize}
\item  it is most probable that most of the cloud, outside the densest regions (i.e. the two PSC and the ridge in between) have an ortho/para H$_2$ ratio also above 5\%, and possibly 10\%, contrarily to what is usually assumed in models.
\item In principle, it should be possible to  fit the PSC profile with this chemical model combined with a dynamical model including depletion, 
to set an age to this PSC and possibly discriminate between several types of collapse but it is beyond the scope of this paper.
\end{itemize}

 We already have some indication that the age of the PSC 
is somewhat above 1.5--2 \pdix{5} years though the N$_2$ abundance suggests a relatively short time (10$^5$ years, except if depletion is compensating for its formation) when we compare our adjusted abundance to the formation rate of N$_2$ given by \citet{Flower06b}. The low abundance of N$_2$ needed to explain the observed \ndhpb abundance indicates that its depletion is not a real problem, though, obviously, \ndhpb would be much more abundant if N$_2$ was not being depleted but CO still was.

Finally, we stress the importance of considering ortho/meta/para chemistry when dealing with the deuteration of the interstellar medium. The importance of 
the ortho--H$_2$ on the amount of deuteration and the observations limited to only ortho--\hddpb make this inclusion compulsory. Moreover, a complete state-to-state chemical model should be developed to take into account rotational pumping, leading to a higher destruction rate of the deuterated trihydrogen cation and possibly explaining the observed linewidth of ortho--\hddp. 

Detecting 
para--\ddhpb would be highly desirable to help constrain the models, but the high frequency and limited transparency of the atmosphere make it a difficult tool to use. 
Though observable from the ground, because of its weakness in cold dark clouds, which are the only places where it should be found, direct para--\ddhpb 
observations should be made on a large number of Galactic lines of sight using the HIFI receiver on board the Heschel Space Observatory.

\begin{acknowledgements}
     We would like to thank D. Flower, G. Pineau des For\^ets, M. Walmsley, S. Cazaux and V. Wakelam for fruitful discussions and an anonymous referee for her/his careful reading.
      Part of this work was supported by the National Science Foundation (NSF) grant AST 05-40882 to the CSO. The authors are grateful to the CSO and JCMT staffs for their support 
      at the telescopes. We would like to thank Samantha Santos for the help provided in numerical calculations of thermal DR rate coefficients. 
      This work has been supported by the NSF under Grants No. PHY-0427460 and PHY-0427376, by an allocation of NERSC and 
      NCSA (project \# PHY-040022) supercomputing resources. This work has benefited from research funding from the European Community's Sixth 
      Framework Programme under RadioNet contract R113CT 2003 5158187
\end{acknowledgements}

\Online

\begin{appendix}
\onecolumn
\section{The reaction rate table used in the Nahoon modified chemical model}
\begin{longtable}[ht]
{lclclclcllcc}
%\caption{Reactions used in the model. The rate coefficient is given at 7K.\label{reactions}}
%\hline
\hline
&&&Reaction       &           &&&   &&    Rate           &  $\alpha$               & Reference \\
   &&&&&&                    &&&                              &cm$^3$~s$^{-1}$  &\\
\hline
\endhead
\endfoot

H$^+$&+&o--\hdb& $\rightarrow$ & H$^+$ & + & p--\hdb & & & Ê& 2.2 10$^{-10}$ & (1) Ê\\
p--\htpb & + & o--\hdb & $\rightarrow$ & p--\htpb & + & p--\hdb & & & Ê& 3.3 10$^{-10}$ & (2) Ê\\%k0$_{popp}\
 Ê& & Ê& $\rightarrow$ & o--\htpb & + & p--\hdb & &&k0$_{poop}$ & 3.9 10$^{-10}$ & (2)\\
 & Ê& Ê& $\rightarrow$ & o--\htpb & + & o--\hdb & &&k0$_{pooo}$ & 7.8 10$^{-12}$ & (2)\\
o--\htpb & + & o--\hdb & $\rightarrow$ & p--\htpb & + & p--\hdb & &&k0$_{oopp}$ & 1.0 10$^{-10}$ & (2)\\
 & Ê& Ê& $\rightarrow$ & p--\htpb & + & o--\hdb & &&k0$_{oopo}$ & 4.1 10$^{-10}$ & Ê(2)\\
 & Ê& Ê& $\rightarrow$ & o--\htpb & + & p--\hdb & &&k0$_{ooop}$ & 1.0 10$^{-10}$ & Ê(2)\\
p--\htpb & + & HD & $\rightarrow$ & o--\htpb & + & HD Ê& &&k1$_{pdod}$ & 8.2 10$^{-13}$ & Ê(2)\\
 & Ê& & $\rightarrow$ & p--\hddp & + & p--\hdb   & & &k1$_{pdpp}$ & 3.2 10$^{-10}$ & Ê(2)\\ %k1$_{pdpp}$
 & Ê& & $\rightarrow$ & p--\hddp & + & o--\hdb   & & &k1$_{pdpo}$ & 4.3 10$^{-10}$ & (2)\\ %k1$_{pdpo}$
 & Ê& & $\rightarrow$ & o--\hddp & + & p--\hdb   & & &k1$_{pdop}$ Ê& 6.8 10$^{-10}$ & Ê(2)\\%k1$_{pdop}$
 & Ê& & $\rightarrow$ & o--\hddp & + & o--\hdb   & & &k1$_{pdoo}$ & 1.6 10$^{-11}$ & Ê(2)\\%k1$_{pdoo}$
o--\htpb & + & HD & $\rightarrow$ & p--\hddp & + & o--\hdb   & &&k1$_{odpo}$ & 1.7 10$^{-10}$ & Ê(2)\\
 & Ê& & $\rightarrow$ & p--\htpb & + & HD Ê& &&k1$_{odpd}$ & 4.35 10$^{-11}$ & Ê(2)\\
 & Ê& & $\rightarrow$ & o--\hddp & + & p--\hdb   & &&k1$_{odop}$ & 2.3 10$^{-10}$ & Ê(2)\\
 & Ê& & $\rightarrow$ & o--\hddp & + & o--\hdb   & &&k1$_{odoo}$ & 1.0 10$^{-9}$ & (2)\\
p--\hddpb & + &o--\hdb Ê& $\rightarrow$ & o--\hddpb & + &p--\hdb   & & & & 1.3 10$^{-9}$ & (2)\\%k-1$_{poop}$
& Ê& Ê& $\rightarrow$ & p--\htpb & + & HD   & &&k\_1$_{popd}$ & 3.9 10$^{-14}$ Ê& (2)\\
& Ê& Ê& $\rightarrow$ & o--\hddpb & + &o--\hdb   & & & & 2.0 10$^{-15}$ Ê& (2)\\%k-1$_{pooo}$
o--\hddpb & + &o--\hdb Ê& $\rightarrow$ & p--\htpb & + &HD   & &&k\_1$_{oopd}$ & 7.7 10$^{-11}$ & (2)\\
 & Ê& Ê& $\rightarrow$ & o--\htpb & + &HD   & &&k\_1$_{oood}$ & 8.3 10$^{-11}$ Ê& (2)\\
 & Ê& Ê& $\rightarrow$ & p--\hddpb & + &p--\hdb   & & & & 9.2 10$^{-11}$ Ê& (2)\\%k-1$_{oopp}$
 & Ê& Ê& $\rightarrow$ & p--\hddpb & + &o--\hdb   & & & & 1.8 10$^{-10}$ Ê& (2)\\%k-1$_{oopo}$
 & Ê& Ê& $\rightarrow$ & o--\hddpb & + &p--\hdb   & & & & 2.35 10$^{-10}$ Ê& (2)\\%k-1$_{ooop}$
o--\hddpb & + &p--\hdb Ê& $\rightarrow$ & p--\hddpb & + &o--\hdb   & & & & 4.2 10$^{-15}$ Ê& (2)\\%k-1$_{oppo}$
o--\hddpb & + & HD& $\rightarrow$ & p--\htpb & + & Êo--\ddb Ê& & & & 1.2 10$^{-15}$   Ê& (2)\\%k2$_{odpo2}$
Ê& Ê& Ê& $\rightarrow$ & p--\hddpb & + & HD   & & & & 5.4 10$^{-11}$ Ê& (2)\\%k2$_{odpd}$
 & Ê& Ê& $\rightarrow$ & p--\ddhpb & + & Êp--\hdb Ê& & & & 7.3 10$^{-11}$   Ê& (2)\\%k2$_{odpp}$
 & Ê& Ê& $\rightarrow$ & p--\ddhpb & + & Êo--\hdb Ê& & & & 3.8 10$^{-10}$   & (2)\\%k2$_{odpo}$
 & Ê& Ê& $\rightarrow$ & o--\ddhpb & + & Êp--\hdb Ê& & & & 1.8 10$^{-10}$   Ê& (2)\\%k2$_{odop}$
 & Ê& Ê& $\rightarrow$ & o--\ddhpb & + & Êo--\hdb Ê& & & & 6.8 10$^{-10}$   Ê& (2)\\%k2$_{odoo}$
 p--\hddpb & + & HD Ê& $\rightarrow$ & p--\ddhpb & + & Êp--\hdb Ê& & & & 4.8 10$^{-10}$   & (2)\\%k2$_{pdpp}$
 & Ê& Ê& $\rightarrow$ & p--\ddhpb & + & Êo--\hdb Ê& & & & 2.1 10$^{-12}$   Ê& (2)\\%k2$_{pdpo}$
 & Ê& Ê& $\rightarrow$ & o--\ddhpb & + & Êp--\hdb Ê& & & & 7.5 10$^{-10}$   Ê& (2)\\%k2$_{pdop}$
 & Ê& Ê& $\rightarrow$ & o--\ddhpb & + & Êo--\hdb Ê& & & & 2.3 10$^{-10}$   Ê& (2)\\%k2$_{pdoo}$
 p--\ddhpb & + &o--\hdb Ê& $\rightarrow$ & p--\hddpb & + & HD   & & & & 3.1 10$^{-11}$ Ê& (2)\\%k-2$_{popd}$
 & Ê& Ê& $\rightarrow$ & p--\ddhpb & + & p--\hdb   & & & & 3.3 10$^{-10}$ Ê& (2)\\%k-2$_{popp}$
 & Ê& Ê& $\rightarrow$ & o--\hddpb & + & HD   & & & & 1.2 10$^{-13}$ Ê& (2)\\%k-2$_{pood}$
o--\ddhpb & + &o--\hdb Ê& $\rightarrow$ & p--\hddpb & + & HD   & & & & 3.8 10$^{-12}$ Ê& (2)\\%k-2$_{oopd}$
 & Ê& Ê& $\rightarrow$ & o--\ddhpb & + & p--\hdb   & & & & 4.45 10$^{-10}$ Ê& (2)\\%k-2$_{ooop}$
p--\ddhpb & + & HD Ê& $\rightarrow$ & Êp--\hddpb & + & Êo--\ddb Ê& & & & 2.8 10$^{-15}$   Ê& (2)\\%k3$_{pdpo}$
 & Ê& Ê& $\rightarrow$ & o--\ddhpb & + & ÊHD Ê& & & & 9.1 10$^{-11}$   Ê& (2)\\%k3$_{pdod}$
 & Ê& Ê& $\rightarrow$ & p--\dtpb & + & Êp--\hdb Ê& & & & 3.3 10$^{-11}$   Ê& (2)\\%k3$_{pdpp}$
 & Ê& Ê& $\rightarrow$ & p--\dtpb & + & Êo--\hdb Ê& & & & 9.75 10$^{-11}$   Ê& (2)\\%k3$_{pdpo}$
 & Ê& Ê& $\rightarrow$ & m-\dtpb & + & Êp--\hdb Ê& & & & 2.6 10$^{-10}$   Ê& (2)\\%k3$_{pdmp}$
 & Ê& Ê& $\rightarrow$ & m-\dtpb & + & Êo--\hdb Ê& & & & 7.8 10$^{-10}$   Ê& (2)\\%k3$_{pdmp}$
 o--\ddhpb & + Ê& HD & $\rightarrow$ & m-\dtpb & + & Êp--\hdb Ê& & & & 1.6 10$^{-10}$   Ê& (2)\\%k3$_{odmp}$
 & Ê& Ê& $\rightarrow$ & m-\dtpb & + & Êo--\hdb Ê& & & & 4.5 10$^{-10}$   Ê& (2)\\%k3$_{odmo}$
 & Ê& Ê& $\rightarrow$ & o--\dtpb & + & Êp--\hdb Ê& & & & 1.9 10$^{-10}$   Ê& (2)\\%k3$_{odop}$
 & Ê& Ê& $\rightarrow$ & o--\dtpb & + & Êo--\hdb Ê& & & & 4.7 10$^{-10}$   Ê& (2)\\%k3$_{odoo}$
 & Ê& Ê& $\rightarrow$ & p--\ddhpb & + & ÊHD Ê& & & & 9.9 10$^{-14}$   Ê& (2)\\%k3$_{odpd}$
m-\dtpb & + &o--\hdb Ê& $\rightarrow$ & o--\ddhpb & + & HD   & & & & 1.3 10$^{-11}$ Ê& (2)\\%k-3$_{mood}$
 & Ê& Ê& $\rightarrow$ & p--\ddhpb & + & HD   & & & & 1.9 10$^{-14}$ Ê& (2)\\%k-3$_{mopd}$
o--\dtpb & + &o--\hdb Ê& $\rightarrow$ & o--\ddhpb & + & HD   & & & & 4.4 10$^{-14}$ Ê& (2)\\%k-3$_{oood}$
p--\dtpb & + &o--\hdb Ê& $\rightarrow$ & p--\ddhpb & + & HD   & & & & 2.6 10$^{-13}$ Ê& (2)\\%k-3$_{popd}$
p--\dtpb & + & HD Ê& $\rightarrow$ & m-\dtpb & + & HD   & & & & 7.75 10$^{-10}$ Ê& (2)\\%k4$_{pdmd}$
m-\dtpb & + & HD Ê& $\rightarrow$ & o--\dtpb & + & HD   & & & & 2.5 10$^{-10}$ Ê& (2)\\%k4$_{mdod}$
 & Ê&   & $\rightarrow$ & p--\dtpb & + & HD   & & & & 9.5 10$^{-12}$ Ê& (2)\\%k4$_{mdpd}$
o--\dtpb & + & HD Ê& $\rightarrow$ & m-\dtpb & + & HD   & & & & 8.45 10$^{-13}$ Ê& (2)\\%k4$_{odmd}$
o--\htpb & + & o--\ddb & $\rightarrow$ & o--\hddpb & + & HD & & & Ê& 2.1 10$^{-10}$ & (2)\\%k5$_{oo2od}$
 & Ê& Ê& $\rightarrow$ & o--\ddhpb & + & o--\hdb & & & Ê& 1.3 10$^{-9}$ & (2)\\%k5$_{oo2oo}$
o--\htpb & + & p--\ddb & $\rightarrow$ & o--\hddpb & + & HD & & & Ê& 5.0 10$^{-10}$ & (2)\\%k5$_{op2od}$
 & Ê& Ê& $\rightarrow$ & p--\ddhpb & + & o--\hdb & & & Ê& 9.7 10$^{-10}$ & (2)\\%k5$_{op2po}$
p--\htpb & + & o--\ddb & $\rightarrow$ & p--\hddpb & + & HD & && Ê& 1.6 10$^{-10}$ & (2)\\%k5$_{po2pd}$
 & Ê& Ê& $\rightarrow$ & o--\hddpb & + & HD & && Ê& 1.4 10$^{-10}$ & (2)\\%k5$_{po2od}$
 & Ê& Ê& $\rightarrow$ & o--\ddhpb & + & p--\hdb & & & & 5.95 10$^{-10}$ & (2)\\%k5$_{po2op}$
 & Ê& Ê& $\rightarrow$ & o--\ddhpb & + & o--\hdb & & & Ê& 6.2 10$^{-10}$ & (2)\\%k5$_{po2oo}$
p--\htpb & + & p--\ddb & $\rightarrow$ & p--\hddpb & + & HD & & & Ê& 2.55 10$^{-10}$ & (2)\\%k5$_{pp2pd}$
 & Ê& Ê& $\rightarrow$ & o--\hddpb & + & HD & & & Ê& 3.45 10$^{-10}$ & (2)\\%k5$_{pp2od}$
 & Ê& Ê& $\rightarrow$ & p--\ddhpb & + & p--\hdb & & & Ê& 4.0 10$^{-10}$ & (2)\\%k5$_{pp2pp}$
 & Ê& Ê& $\rightarrow$ & p--\ddhpb & + & o--\hdb & & & Ê& 4.9 10$^{-10}$ & (2)\\%k5$_{pp2po}$
o--\hddpb & + & o--\ddb & $\rightarrow$ & p--\ddhpb & + & HD & & & Ê& 1.7 10$^{-10}$ & (2)\\%k6$_{oo2pd}$
 & Ê& Ê& $\rightarrow$ & o--\ddhpb & + & HD & && Ê& 3.9 10$^{-10}$ & (2)\\%k6$_{oo2od}$
 & Ê& Ê& $\rightarrow$ & m-\dtpb & + & o--\hdb & & & Ê& 3.2 10$^{-10}$ & (2)\\%k6$_{oo2mo}$
 & Ê& Ê& $\rightarrow$ & o--\dtpb & + & o--\hdb & & & Ê& 5.3 10$^{-10}$ & (2)\\%k6$_{oo2oo}$
o--\hddpb & + & p--\ddb & $\rightarrow$ & o--\hddpb & + & o--\ddb & & &Ê& 4.8 10$^{-11}$ & (2)\\%k6$_{op2oo2}$
 & Ê& Ê& $\rightarrow$ & p--\ddhpb & + & HD & & & Ê& 4.1 10$^{-10}$ & (2)\\%k6$_{op2pd}$ 
 & Ê& Ê& $\rightarrow$ & o--\ddhpb & + & HD & & & Ê& 3.6 10$^{-10}$ & (2)\\%k6$_{op2od}$
 & Ê& Ê& $\rightarrow$ & p--\dtpb & + & o--\hdb & & & Ê& 6.1 10$^{-11}$ & (2)\\%k6$_{op2po}$
 & Ê& Ê& $\rightarrow$ & m-\dtpb & + & o--\hdb & & & Ê& 5.05 10$^{-10}$ & (2)\\%k6$_{op2mo}$
p--\hddpb & + & o--\ddb & $\rightarrow$ & p--\ddhpb & + & HD & & & Ê& 1.2 10$^{-10}$ & (2)\\%k6$_{po2pd}$
 & Ê& Ê& $\rightarrow$ & o--\ddhpb & + & HD & & & Ê& 2.7 10$^{-10}$ & (2)\\%k6$_{po2od}$
 & Ê& Ê& $\rightarrow$ & m-\dtpb & + & p--\hdb & & & Ê& 4.3 10$^{-10}$ & (2)\\%k6$_{po2mp}$
 & Ê& Ê& $\rightarrow$ & o--\dtpb & + & p--\hdb & && Ê& 5.8 10$^{-10}$ & (2)\\%k6$_{po2op}$
p--\hddpb & + & p--\ddb & $\rightarrow$ & p--\hddpb & + & o--\ddb & && Ê& 6.55 10$^{-11}$ & (2)\\%k6$_{pp2po2}$
 & Ê& Ê& $\rightarrow$ & p--\ddhpb & + & HD & && Ê& 4.1 10$^{-10}$ & (2)\\%k6$_{pp2pd}$
 & Ê& Ê& $\rightarrow$ & o--\ddhpb & + & HD & && Ê& 2.8 10$^{-10}$ & (2)\\%k6$_{pp2od}$
 & Ê& Ê& $\rightarrow$ & p--\dtpb & + & p--\hdb & && Ê& 6.9 10$^{-11}$ & (2)\\%k6$_{pp2pp}$
 & Ê& Ê& $\rightarrow$ & m-\dtpb & + & p--\hdb & && Ê& 5.8 10$^{-10}$ & (2)\\%k6$_{pp2mp}$
o--\ddhpb & + & o--\ddb & $\rightarrow$ & p--\ddhpb & + & o--\ddb & && Ê& 3.85 10$^{-14}$ & (2)\\%k7$_{oo2po2}$
 & Ê& Ê& $\rightarrow$ & m-\dtpb & + & HD & &&& 5.7 10$^{-10}$ & (2)\\%k7$_{oo2md}$
 & Ê& Ê& $\rightarrow$ & o--\dtpb & + & HD & &&Ê& 6.4 10$^{-10}$ & (2)\\%k7$_{oo2od}$
o--\ddhpb & + & p--\ddb & $\rightarrow$ & p--\ddhpb & + & p--\ddb & && Ê& 2.1 10$^{-14}$ & (2)\\%k7$_{op2pp2}$
 & Ê& Ê& $\rightarrow$ & p--\ddhpb & + & o--\ddb & && Ê& 1.6 10$^{-10}$ & (2)\\%k7$_{op2po2}$ 
 & Ê& Ê& $\rightarrow$ & o--\ddhpb & + & o--\ddb & && Ê& 8.8 10$^{-11}$ & (2)\\%k7$_{op2oo2}$ 
 & Ê& Ê& $\rightarrow$ & p--\dtpb & + & HD & && & 8.4 10$^{-11}$ & (2)\\%k7$_{op2pd}$
 & Ê& Ê& $\rightarrow$ & m-\dtpb & + & HD & && Ê& 6.4 10$^{-10}$ & (2)\\%k7$_{op2md}$
 & Ê& Ê& $\rightarrow$ & o--\dtpb & + & HD & && Ê& 3.5 10$^{-10}$ & (2)\\%k7$_{op2od}$
p--\ddhpb & + & o--\ddb & $\rightarrow$ & o--\ddhpb & + & p--\ddb & && Ê& 5.0 10$^{-13}$ & (2)\\%k7$_{po2op2}$
 & Ê& Ê& $\rightarrow$ & o--\ddhpb & + & o--\ddb & && Ê& 3.7 10$^{-11}$ & (2)\\%k7$_{po2oo2}$ 
 & Ê& Ê& $\rightarrow$ & p--\dtpb & + & HD & && Ê& 7.7 10$^{-11}$ & (2)\\%k7$_{po2pd}$
 & Ê& Ê& $\rightarrow$ & m-\dtpb & + & HD & && Ê& 8.6 10$^{-10}$ & (2)\\%k7$_{po2md}$ 
 & Ê& Ê& $\rightarrow$ & o--\dtpb & + & HD & && Ê& 3.05 10$^{-10}$ & (2)\\%k7$_{po2od}$ 
p--\ddhpb & + & p--\ddb & $\rightarrow$ & p--\ddhpb & + & o--\ddb & && Ê& 3.7 10$^{-11}$ & (2)\\%k7$_{pp2po}$
 & Ê& Ê& $\rightarrow$ & o--\ddhpb & + & p--\ddb & && Ê& 2.0 10$^{-11}$ & (2)\\%k7$_{pp2op2}$
 & Ê& Ê& $\rightarrow$ & o--\ddhpb & + & o--\ddb & && Ê& 8.5 10$^{-11}$ & (2)\\%k7$_{pp2oo2}$
 & Ê& Ê& $\rightarrow$ & p--\dtpb & + & HD & && Ê& 1.15 10$^{-10}$ & (2)\\%k7$_{pp2pd}$
 & Ê& Ê& $\rightarrow$ & m-\dtpb & + & HD & && Ê& 9.5 10$^{-10}$ & (2)\\%k7$_{pp2md}$
m-\dtpb & + & o--\ddb & $\rightarrow$ & p--\dtpb & + & o--\ddb & && Ê& 8.6 10$^{-12}$ & (2)\\%k8$_{mo2po2}$
 & Ê& Ê& $\rightarrow$ & m-\dtpb & + & p--\ddb & && Ê& 1.7 10$^{-15}$ & (2)\\%k8$_{mo2mp2}$
 & Ê& Ê& $\rightarrow$ & o--\dtpb & + & p--\ddb & && Ê& 1.2 10$^{-12}$ & (2)\\%k8$_{mo2op2}$
 & Ê& Ê& $\rightarrow$ & o--\dtpb & + & o--\ddb & && Ê& 3.0 10$^{-10}$ & (2)\\%k8$_{mo2oo2}$ 
m-\dtpb & + & p--\ddb & $\rightarrow$ & p--\dtpb & + & p--\ddb & && Ê& 5.9 10$^{-12}$ & (2)\\%k8$_{mp2pp2}$
 & Ê& Ê& $\rightarrow$ & p--\dtpb & + & o--\ddb & && Ê& 4.7 10$^{-11}$ & (2)\\%k8$_{mp2po2}$
 & Ê& Ê& $\rightarrow$ & m-\dtpb & + & o--\ddb & && Ê& 6.1 10$^{-10}$ & (2)\\%k8$_{mp2mo2}$ 
 & Ê& Ê& $\rightarrow$ & o--\dtpb & + & p--\ddb & && Ê& 4.7 10$^{-11}$ & (2)\\%k8$_{mp2op2}$
 & Ê& Ê& $\rightarrow$ & o--\dtpb & + & o--\ddb & && Ê& 5.1 10$^{-11}$ & (2)\\%k8$_{mp2oo2}$
o--\dtpb & + & o--\ddb & $\rightarrow$ & m-\dtpb & + & o--\ddb & && Ê& 1.0 10$^{-12}$ & (2)\\%k8$_{oo2mo2}$
 & Ê& Ê& $\rightarrow$ & o--\dtpb & + & p--\ddb & && Ê& 1.15 10$^{-15}$ & (2)\\%k8$_{oo2op2}$
o--\dtpb & + & p--\ddb & $\rightarrow$ & p--\dtpb & + & o--\ddb & && Ê& 9.2 10$^{-11}$ & (2)\\%k8$_{op2po2}$
 & Ê& Ê& $\rightarrow$ & m-\dtpb & + & p--\ddb & && Ê& 1.6 10$^{-13}$ & (2)\\%k8$_{op2mp2}$
 & Ê& Ê& $\rightarrow$ & m-\dtpb & + & o--\ddb & && Ê& 5.7 10$^{-10}$ & (2)\\%k8$_{op2mo2}$
 & Ê& Ê& $\rightarrow$ & o--\dtpb & + & o--\ddb & && Ê& 1.6 10$^{-10}$ & (2)\\%k8$_{op2oo2}$
 p--\dtpb & + & o--\ddb & $\rightarrow$ & m-\dtpb & + & p--\ddb & && Ê& 2.8 10$^{-14}$ & (2)\\%k8$_{po2mp2}$ 
 & Ê& Ê& $\rightarrow$ & m-\dtpb & + & o--\ddb & && Ê& 7.0 10$^{-10}$ & (2)\\%k8$_{po2mo2}$
 & Ê& Ê& $\rightarrow$ & o--\dtpb & + & p--\ddb & && Ê& 1.7 10$^{-11}$ & (2)\\%k8$_{po2op2}$
p--\dtpb & + & p--\ddb & $\rightarrow$ & m-\dtpb & + & p--\ddb & && Ê& 4.8 10$^{-10}$ & (2)\\%k8$_{pp2mp2}$
 & Ê& Ê& $\rightarrow$ & m-\dtpb & + & o--\ddb & && Ê& 4.4 10$^{-10}$ & (2)\\%k8$_{pp2mo2}$
o--H$_3^+$ & + & CO & $\rightarrow$ & HCO$^+$ Ê& + & o--\hdb     Ê&&& k$_{CO}$   & 1.6 10$^{-9}$   & (3)\\
p--H$_3^+$ & + & CO & $\rightarrow$ & HCO$^+$ Ê& + & p--\hdb     Ê&&& 1/2 k$_{CO}$   & 8.1 10$^{-10}$   & (3)\\
&& & $\rightarrow$ & HCO$^+$ & + & o--\hdb Ê&& & 1/2 k$_{CO}$   & 8.1 10$^{-10}$   & (3)\\
o--\hddpb & + & CO & $\rightarrow$ & HCO$^+$ & + & HD   Ê&& Ê& 2/3 k$'_{CO}$\footnotemark[1]   & 9.4 10$^{-10}$   & (3)\\
&& & $\rightarrow$ & DCO$^+$ & + & o--\hdb   &&& 1/3 k$'_{CO}$   & 4.7 10$^{-10}$   & (3)\\
p--\hddpb & + & CO & $\rightarrow$ & HCO$^+$ & + & HD     Ê&&& 2/3 k$'_{CO}$   & 9.4 10$^{-10}$   & (3)\\
&& & $\rightarrow$ & DCO$^+$ & + & p--\hdb   Ê&&& 1/3 k$'_{CO}$   & 4.7 10$^{-10}$   & (3)\\
o--\ddhpb & + & CO & $\rightarrow$ & HCO$^+$ & + & o--D$_2$   Ê&& Ê& 1/3 k$''_{CO}$   & 4.3 10$^{-10}$   & (3)\\
&& & $\rightarrow$ & DCO$^+$ & + & HD   Ê&&& 2/3 k$''_{CO}$   & 8.6 10$^{-10}$   & (3)\\
p--\ddhpb & + & CO & $\rightarrow$ & HCO$^+$ & + & p--D$_2$     Ê&& & $1/3$k$''_{CO}$   & 4.3 10$^{-10}$   & (3)\\
&& & $\rightarrow$ & DCO$^+$ & + & HD   &&& 2/3 k$''_{CO}$   & 8.6 10$^{-10}$   & (3)\\
o--\dtpb & + & CO & $\rightarrow$ & DCO$^+$ & + & o--D$_2$     Ê& &&k$'''_{CO}$   & 1.2 10$^{-9}$   & (3)\\
m-\dtpb & + & CO & $\rightarrow$ & DCO$^+$ & + & o--D$_2$     Ê&&& 1/2 k$'''_{CO}$   & 6.0 10$^{-10}$   & (3)\\
&& & $\rightarrow$ & DCO$^+$ & + & p--D$_2$     && & 1/2 k$'''_{CO}$   & 6.0 10$^{-10}$   & (3)\\
p--\dtpb & + & CO & $\rightarrow$ & DCO$^+$ & + & p--D$_2$     Ê& &&k$'''_{CO}$   & 1.2 10$^{-9}$   & (3)\\
p--\htpb & + & N$_2$ & $\rightarrow$ & N$_2$H$^+$ & + & o-H$_2$ Ê&&&   1/2 k$_{N_2}$ & 8.5 10$^{-10}$ & (3)\\
&& & $\rightarrow$ & ÊN$_2$H$^+$ & + & p-H$_2$ Ê&   && 1/2 k$_{N_2}$ & 8.5 10$^{-10}$ & (3)\\
o--\htpb & + & N$_2$ & $\rightarrow$ & N$_2$H$^+$ & + & o-H$_2$ Ê&   &&k$_{N_2}$ & 1.7 10$^{-9}$ & (3)\\
o--\hddpb & + & N$_2$ & $\rightarrow$ & N$_2$H$^+$ & + & HD Ê&&& Ê2/3 k$'_{N_2}$\footnotemark[1]& 1.0 10$^{-9}$ & (3)\\
\footnotetext[1]{k$'_{CO}$ and k$'_{N_2}$ indicate that the k$_{CO}$ and k$_{N_2}$ rates have been corrected for the higher reduced mass of the system because of the presence of \hddpb instead of \htp. Similarly, k$''$ and k$'''$ are given for \ddhpb  and \dtpb isotopologues.}
&& & $\rightarrow$ & N$_2$D$^+$ & + & o-H$_2$ && & Ê1/3 k$'_{N_2}$ & 5.0 10$^{-10}$ & (3)\\
p--\hddpb & + & N$_2$ & $\rightarrow$ & N$_2$H$^+$ & + & HD Ê&&& Ê2/3 k$'_{N_2}$ & 1.0 10$^{-9}$ & (3)\\
&& & $\rightarrow$ & N$_2$D$^+$ & + & p-H$_2$ Ê&&& Ê1/3 k$'_{N_2}$ & 5.0 10$^{-10}$ & (3)\\
o--\ddhpb & + & N$_2$ & $\rightarrow$ & N$_2$H$^+$ & + & p--D$_2$ Ê&&& Ê1/3 k$''_{N_2}$ & 4.5 10$^{-10}$ & (3)\\
&& & $\rightarrow$ & N$_2$D$^+$ & + & HD Ê&&& Ê2/3 k$''_{N_2}$ & 9.1 10$^{-10}$ & (3)\\
p--\ddhpb & + & N$_2$ & $\rightarrow$ & N$_2$H$^+$ & + & p--D$_2$ Ê&&& Ê1/3 k$''_{N_2}$ & 4.5 10$^{-10}$ & (3)\\
&& & $\rightarrow$ & N$_2$D$^+$ & + & HD Ê& Ê&& 2/3 k$''_{N_2}$ & 9.1 10$^{-10}$ & (3)\\
o--\dtpb & + & N$_2$ & $\rightarrow$ & N$_2$D$^+$ & + & o--D$_2$ Ê& &&k$'''_{N_2}$ & 1.3 10$^{-9}$ & (3)\\
m-\dtpb & + & N$_2$ & $\rightarrow$ & N$_2$D$^+$ & + & o--D$_2$ Ê&&& 1/2 k$'''_{N_2}$ & 6.3 10$^{-10}$ & (3)\\
&& & $\rightarrow$ & N$_2$D$^+$ & + & p--D$_2$ Ê&&& 1/2 k$'''_{N_2}$ & 6.3 10$^{-10}$ & (3)\\
p--\dtpb & + & N$_2$ & $\rightarrow$ & N$_2$D$^+$ & + & p--D$_2$ Ê& &&k$'''_{N_2}$ & 1.3 10$^{-9}$ & (3)\\
H$^+$ & + & e$^-$ & $\rightarrow$ &   H&&& &&k$_{rec}$ Ê& Ê4.9 10$^{-11}$& (3) \\
o--H$_3^+$ & + & e$^-$ & $\rightarrow$ & H& + &H &+& H &3/4 o--k$_{rec1}$ &   3.6 10$^{-8}$ & (4,5) \\
&&& $\rightarrow$ & o--\hdb& + &H & &&1/4 o--k$_{rec1}$ &   1.2 10$^{-8}$ & Ê(4,5)\\
p--H$_3$$^+$ & + & e$^-$ & $\rightarrow$ & ÊH& + &H &+& H&3/4 p--k$_{rec1}$ &   5.6 10$^{-7}$ & (4,5) \\
&&& $\rightarrow$ & Êp--\hdb& + &H & &&1/8 p--k$_{rec1}$ &   9.4 10$^{-8}$ & (4,5) \\
&& & $\rightarrow$ & Êo--\hdb& + &H & &&1/8 p--k$_{rec1}$ &   9.4 10$^{-8}$ & (4,5) \\
o--H$_2$D$^+$ & + & e$^-$ & $\rightarrow$ & H& + &H& +& D&3/4 o--k$_{rec2}$ & 2.6 10$^{-7}$ &(4,6) \\
&& & $\rightarrow$ & o--\hdb& + &D & &&$\sim$0.07 o--k$_{rec2}$ & 2.5 10$^{-8}$ & (4,6)\\
&& & $\rightarrow$ & HD& + &D & &&$\sim$0.18 o--k$_{rec2}$ & 6.25 10$^{-8}$ &(4,6) \\
p--H$_2$D$^+$ & + & e$^-$ & $\rightarrow$ & H& + &H& +& D &3/4 p--k$_{rec2}$ & 1.8 10$^{-7}$ &(4,6) \\
&&& $\rightarrow$ & p--\hdb& + &D & &&$\sim$0.07 p--k$_{rec2}$ & 1.7 10$^{-8}$ &(4,6) \\
&&& $\rightarrow$ & HD& + &D & &&$\sim$0.18 p--k$_{rec2}$ & 4.3 10$^{-8}$ &(4,6) \\
o--D$_2$H$^+$ & + & e$^-$ & $\rightarrow$ & H& + &D& + &D &$\sim$0.77 o--k$_{rec3}$   & Ê4.3 10$^{-8}$ Ê&(4,7) \\
&&& $\rightarrow$ & HD& + &D & &&$\sim$0.13 o--k$_{rec3}$ & 7.3 10$^{-9}$ & (4,7)\\
&&& $\rightarrow$ & o--D$_2$& + &H & &&$\sim$0.10 o--k$_{rec3}$ & 5.6 10$^{-9}$ &(4,7) \\
p--D$_2$H$^+$ & + & e$^-$ & $\rightarrow$ & H& + &D& +& D&$\sim$0.77 p--k$_{rec3}$   & Ê4.6 10$^{-8}$ Ê& (4,7)\\
&& & $\rightarrow$ & HD& + &D & &&$\sim$0.13 p--k$_{rec3}$ & 7.8 10$^{-9}$ &(4,7) \\
&& & $\rightarrow$ & p--D$_2$& + &H & &&$\sim$0.10 p--k$_{rec3}$ & 6.0 10$^{-9}$ &(4,7) \\
o--D$_3$$^+$ & + & e$^-$ & $\rightarrow$ & D& + &D &+& D&3/4 o--k$_{rec4}$ Ê& 1.8 10$^{-7}$ & (4,8) \\
&&& $\rightarrow$ & o--D$_2$& + &D & &&1/4 o--k$_{rec4}$ & 6.1 10$^{-8}$ &(4,8) \\
p--D$_3$$^+$ & + & e$^-$ & $\rightarrow$ & ÊD& + &D &+& D&3/4 p--k$_{rec4}$ Ê& 1.5 10$^{-7}$ & (4,8) \\
&& & $\rightarrow$ & Êp--D$_2$ & + &D & &&1/4 p--k$_{rec4}$ Ê& 5.0 10$^{-8}$ &(4,8) Ê\\
m--D$_3$$^+$ & + & e$^-$ & $\rightarrow$ & ÊD& + &D& +& D&3/4 m--k$_{rec4}$ Ê& 3.6 10$^{-7}$ &(4,8) Ê\\
&& & $\rightarrow$ & Êo--D$_2$ & + &D & &&1/8 m--k$_{rec4}$ Ê& 6.0 10$^{-8}$ & (4,8) \\
&& & $\rightarrow$ & Êp--D$_2$ & + &D & &&1/8 m--k$_{rec4}$ Ê& 6.0 10$^{-8}$ & (4,8) \\
N$_2$H$^+$ Ê& + & e$^-$ & $\rightarrow$ & ÊH & + & N$_2$ & Ê&   & k$_{N_2H^+}$ & 6.8 10$^{-7}$ Ê& (9)\\
N$_2$D$^+$ Ê& + & e$^-$ & $\rightarrow$ & ÊD & + & N$_2$ & Ê&   & k$_{N_2D^+}$ & Ê6.8 10$^{-7}$ & (9)\\
HCO$^+$ Ê& + & e$^-$ & $\rightarrow$ & ÊH & + & CO & Ê&   & k$_{HCO^+}$ & 3.2 10$^{-6}$   & (2)\\
DCO$^+$ Ê& + & e$^-$ & $\rightarrow$ & ÊD & + & CO & Ê& Ê& k$_{DCO^+}$ & Ê3.2 10$^{-6}$ & (2)\\
gr$^0$ Ê& + & e$^-$ & $\rightarrow$&   gr$^-$ Ê& Ê& Ê& Ê&   & k$_e$ &   3.7 10$^{-3}$ & (4)\\
H$^+$ & + & gr$^-$ & $\rightarrow$ & ÊH & + & gr$^0$& && k$_{gr}$ Ê& 6.0 10$^{-4}$   & Ê(4)\\
HD$^+$ & + & gr$^-$ & $\rightarrow$ & ÊH & + & D & + & gr$^0$& Ê2/3 k$_{gr1}$ & Ê2.3 10$^{-4}$ Ê& Ê(4)\\
&& &                                 Ê$\rightarrow$ & ÊHD & + & gr$^0$& & & 1/3 k$_{gr1}$ Ê& Ê1.2 10$^{-4}$ Ê& Ê(4)\\
o--D$_2^+$ & + & gr$^-$ & $\rightarrow$ & ÊD & + & D & + & gr$^0$& 2/3 k$_{gr2}$ Ê& 2.0 10$^{-4}$   & Ê(4)\\
 && &                                     $\rightarrow$ & Êo--D$_2$ & + Ê& gr$^0$& & &1/3 k$_{gr2}$  Ê& 1.0 10$^{-4}$   & Ê(4)\\
p--D$_2^+$ & + & gr$^-$ & $\rightarrow$ & ÊD & + & D & + & gr$^0$& 2/3 k$_{gr2}$Ê& Ê2.0 10$^{-4}$ Ê& Ê(4)\\
 Ê&& &                                   Ê$\rightarrow$ & Êp--D$_2$ & + & gr$^0$& & &1/3 k$_{gr2}$ Ê& 1.0 10$^{-4}$   & Ê(4)\\
He$^+$ & + & gr$^-$ & $\rightarrow$ & ÊHe & + & gr$^0$& && Ê& 3.0 10$^{-4}$   & Ê(4)\\
p--H$_3^+$ & + & gr$^-$ & $\rightarrow$ & Êp-H$_2$ & + & H & + & gr$^0$& 1/6 k$_{gr3}$ & Ê5.8 10$^{-5}$ Ê& (4)\\
 && &                                     $\rightarrow$ & Êo-H$_2$ & + & H & + & gr$^0$& 1/6 k$_{gr3}$ & 5.810$^{-5}$   & Ê(4)\\
   && &                                   $\rightarrow$ & Ê3H & + & gr$^0$& + &   & 2/3 k$_{gr3}$ Ê& 2.3 10$^{-4}$ Ê& Ê(4)\\
o--H$_3^+$ & + & gr$^-$ & $\rightarrow$ & Êo-H$_2$ & + & H & + & gr$^0$& 1/3 k$_{gr3}$ & 1.2 10$^{-4}$   & (4)\\
   && &                                   $\rightarrow$ & Ê3H & + & gr$^0$& + &   & 2/3 k$_{gr3}$ Ê& Ê2.3 10$^{-4}$ Ê& Ê(4)\\
p--H$_2$D$^+$ & + & gr$^-$ & $\rightarrow$ & Êp-H$_2$ & + & D & + & gr$^0$& 1/6 k$_{gr4}$ & 5.0 10$^{-5}$   & (4)\\
 && &                                     $\rightarrow$ & ÊHD & + & H & + & gr$^0$& 1/3 k$_{gr4}$ Ê& Ê1.0 10$^{-4}$ Ê& Ê(4)\\
 && &                                     $\rightarrow$ & Ê2H & + & D & + & gr$^0$& 1/2 k$_{gr4}$ & 1.5 10$^{-4}$   & Ê(4)\\
o--H$_2$D$^+$ & + & gr$^-$ & $\rightarrow$ & Ê2H & + & D & + & gr$^0$& 1/2 k$_{gr4}$ & 1.5 10$^{-4}$   & (4)\\
 && &                                     $\rightarrow$ & Êo-H$_2$ & + & D & + & gr$^0$& 1/6 k$_{gr4}$ & 5.0 10$^{-5}$ Ê& Ê(4)\\
&& &                                     Ê$\rightarrow$ & ÊHD & + & H & + & gr$^0$& 1/3 k$_{gr4}$ & Ê1.0 10$^{-4}$ Ê& Ê(4)\\
o--D$_2$H$^+$ & + & gr$^-$ & $\rightarrow$ & Ê2D & + & H & + & gr$^0$& 1/2 k$_{gr5}$ & 1.3 10$^{-4}$   & (4)\\
&& &                                     Ê$\rightarrow$ & Êo--D$_2$ & + & H & + & gr$^0$& 1/6 k$_{gr5}$ & 4.4 10$^{-5}$   & Ê(4)\\
&& &                                     Ê$\rightarrow$ & ÊHD & + & D & + & gr$^0$&1/3 k$_{gr5}$ & 8.9 10$^{-5}$   Ê& Ê(4)\\
p--D$_2$H$^+$ & + & gr$^-$ & $\rightarrow$ & Êp--D$_2$ & + & H & + & gr$^0$& 1/6 k$_{gr5}$& 4.4 10$^{-5}$   Ê& (4)\\
&& &                                     Ê$\rightarrow$ & ÊHD & + & D & + & gr$^0$& 1/3 k$_{gr5}$ & 8.9 10$^{-5}$   Ê& Ê(4)\\
&& &                                     Ê$\rightarrow$ & Ê2D & + & H & + & gr$^0$& 1/2 k$_{gr5}$ & Ê1.3 10$^{-4}$ Ê& Ê(4)\\
o--D$_3^+$ & + & gr$^-$ & $\rightarrow$ & Ê3D & + & gr$^0$& Ê& Ê&2/3 k$_{gr6}$ & 1.6 10$^{-4}$   & (4)\\
 && &                                     $\rightarrow$ & Êo--D$_2$ & + & D & + & gr$^0$& 1/3 k$_{gr6}$ & Ê8.1 10$^{-5}$ Ê& Ê(4)\\
m--D$_3^+$ & + & gr$^-$ & $\rightarrow$ & Ê3D & + & gr$^0$& Ê& Ê& 2/3 k$_{gr6}$ & 1.2 10$^{-4}$   & (4)\\
&& &                                     Ê$\rightarrow$ & Êo--D$_2$ & + & D & + & gr$^0$& 1/6 k$_{gr6}$ & Ê4.1 10$^{-5}$ Ê& Ê(4)\\
 && &                                     $\rightarrow$ & Êp--D$_2$ & + & D & + & gr$^0$& 1/6 k$_{gr6}$Ê& 4.1 10$^{-5}$   & Ê(4)\\
p--D$_3^+$ & + & gr$^-$ & $\rightarrow$ & Ê3D & + & gr$^0$& Ê& Ê& 2/3 k$_{gr6}$& Ê1.6 10$^{-4}$ Ê& (4)\\
 && &                                     $\rightarrow$ & Êp--D$_2$ & + & D & + & gr$^0$& 1/3 k$_{gr6}$Ê& 8.1 10$^{-5}$   & Ê(4)\\
HCO$^+$ Ê& + & gr$^-$ & $\rightarrow$ & ÊH & + & CO & + & gr$^0$& k$_{HCO^+}$ & Ê1.1 10$^{-4}$ Ê& (4)\\
DCO$^+$ Ê& + & gr$^-$ & $\rightarrow$ & ÊD & + & CO & + & gr$^0$& k$_{DCO^+}$ & 1.1 10$^{-4}$ & (4)\\
N$_2$H$^+$ Ê& + & gr$^-$ & $\rightarrow$ & ÊH & + & N$_2$ & + & gr$^0$& k$_{N_2H^+}$ & 1.1 10$^{-4}$  & (4)\\
N$_2$D$^+$ Ê& + & gr$^-$ & $\rightarrow$ & ÊD & + & N$_2$ & + & gr$^0$& k$_{N_2D^+}$ & 1.1 10$^{-4}$Ê& (4)\\

\hline
\caption{Reactions used in the model. The rate coefficients are given for 7K. Reaction rates less than 10$^{-15}$ \cc s$^{-1}$ are not taken into account 
in our models. Reference (1) corresponds to \citet{Gerlich90}, references (2), (3), and (4) correspond to E. Hugo, OSU 07, and this paper respectively. 
For OSU 07, branching ratios involving spin states have been infered from quantum mechanical rules. For reactions involving grains, a grain radius of 0.1 $\mu$m and a sticking coefficient of 1 have been considered. (5) \citet{Datz95b} (6) \citet{Datz95a}, 
(7) \citet{Zhaunerchyk08}, (8) \citet{Larsson97}, (9) \citet{Molek07}.\label{reactions}}
%$^\mathrm{a}$k$'_{CO}$ and k$'_{N_2}$ indicate that the k$_{CO}$ and k$_{N_2}$ rates have been corrected for the higher reduced mass of the system because of the presence of \hddpb instead of \htp. Similarly, k$''$ and k$'''$ are given for \ddhpb  and \dtpb isotopologues.
\end{longtable}
\end{appendix}
\twocolumn
\begin{appendix}
\section{On the rate coefficients for dissociation recombination of H$_3^+$ and its isotopologues}
\label{VK}

The dissociative recombination (DR) rate coefficients for ortho-- and para--H$_3^+$ have been published by \citet{santos07}. Here, we present the results obtained for all four H$_3^+$ isotopologues. The DR rate coefficients for different species of the nuclear spin are calculated using the approach described in a series of papers devoted to DR theory for triatomic molecular ions. See \citet{kokoouline03a,kokoouline03b,santos07} for H$_3^+$ and D$_3^+$ calculations and \citet{kokoouline04b,kokoouline05} for H$_2$D$^+$ and D$_2$H$^+$.  The scope of this paper does not allow us to review the theoretical approach in detail. We only list its main ingredients. 

The theoretical approach is fully quantum mechanical and incorporates no adjustable parameters. It relies on {\it ab initio} calculations of potential surfaces for the ground electronic state of the H$_3^+$ ion and several excited  states of the neutral molecule H$_3$, performed by \citet{mistrik01}.

The total wave function of the system is constructed by an appropriate symmetrization of products of vibrational, rotational, electronic, and nuclear spin factors. Therefore, rovibronic and nuclear spin degrees of freedom are explicitly taken into account.

The electronic Born-Oppenheimer potentials for the four H$_3^+$ (and H$_3$) isotopologues have the $C_{3v}$ symmetry group. The $C_{3v}$ symmetry group has a two--dimensional irreducible representation $E$. The ion has a closed electronic shell. The lowest electronic state of the outer electron in H$_3$ has $p$-wave character. The $p$-wave state of the electron also belongs to the $E$ representation. Due to the Jahn-Teller theorem \citep{landau3}, this leads to a strong non-adiabatic coupling between the $E$-degenerate vibrational modes of the ion and the $p$-wave states of the incident electron. The coupling is responsible for the fast DR rate \citep{kokoouline01} in H$_3^+$. In the present model, only the $p$-wave electronic states are included because other partial waves have a much smaller effect on the DR probability\,: the $s$-wave states have no $E$-type character and, therefore, are only weakly coupled to the dissociative electronic states of H$_3$; $d$-wave electronic states are coupled to the $E$-vibrational modes, but the coupling is rather small because the $d$-wave of the incident electron does not penetrate sufficiently close to the ionic core owing to the $d$-wave centrifugal potential barrier.

All three internal vibrational coordinates are taken into account. Vibrational dynamics of the ionic core are described using the hyper-spherical coordinates, which represent the three vibrational degrees of freedom by a hyperradius and two hyperangles. The hyperradius is treated as a dissociation coordinate that represents uniformly the two possible DR channels, the three-body (such as H+H+H) and two-body (such as H$_2$+H). Although the initial vibrational state of the ion is the ground state, after recombination with the electron, other vibrational states of the ionic target molecule can be populated. Therefore, in general, many vibrational states have to be included in the treatment. In particular, the states of the vibrational continuum have to be included, because only such states can lead to the dissociation of the neutral molecule. The vibrational states of the continuum are obtained using a complex absorbing potential placed at a large hyperradius to absorb the flux of the outgoing dissociative wave. 

Since the rovibrational symmetry is $D_{3h}$ for H$_3^+$ and D$_3^+$ and  $C_{2v}$ for H$_2$D$^+$ and D$_2$H$^+$, the rovibrational functions are classified according to the irreducible representations of the corresponding symmetry groups, i.e. $A_1'$, $A_1''$, $A_2'$, $A_2''$, $E'$, and $E''$ for  $D_{3h}$ and $A_1$, $A_2$, $B_1$, and $B_2$ for  $C_{2v}$.  We use the rigid rotor approximation, i.e. the vibrational and rotational parts of the total wave function are calculated independently by diagonalizing the corresponding Hamiltonians. In our approach, the rotational wave functions must be obtained separately for the ions and for the neutral molecules. They are constructed in a different way for the $D_{3h}$  and $C_{2v}$ cases. The rotational eigenstates and eigenenergies of the $D_{3h}$ molecules are symmetric top wave functions \citep[see, for example][]{bunker98}. They can be obtained analytically if the rotational constants are known. The rotational constants are obtained numerically from vibrational wave functions, i.e. they are calculated separately  for each vibrational level of the target molecule. The rotational functions for the $C_{2v}$ ions are obtained numerically by diagonalizing the asymmetric top Hamiltonian \citep{bunker98,kokoouline05}.

Once the rovibrational wave functions are calculated, we construct the electron-ion scattering matrix ($S$-matrix). The $S$-matrix is calculated in the framework of quantum defect theory (QDT) \citep[see, for example,][]{chang72,kokoouline03b,kokoouline05} using the quantum defect parameters obtained from the {\it ab initio} calculation \citep{mistrik01}. The constructed scattering matrix accounts for the Jahn-Teller effect and is diagonal with respect to the different irreducible representations  $\Gamma$  and the total angular momentum $N$ of the neutral molecule. Thus, the actual calculations are made separately for each $\Gamma$ and $N$. Elements of the matrix describe the scattering amplitudes for the change of the rovibrational state of the ion after a collision with the electron. However, the $S$-matrix is not unitary due to the presence of the dissociative vibrational channels (i.e. continuum vibrational states of the ion, discussed above), which are not explicitly listed in the computed $S$-matrix. The "defect" from unitarity of each column of this $S$-matrix is associated with the dissociation probability of the neutral molecule formed during the scattering process. The dissociation probability per collision is then used to calculate the DR cross-sections and rate coefficients.

The nuclear spin states are characterized by one of the $A_1$,  $A_2$, or $E$ irreducible representations of the symmetry group $S_3$ for $D_{3h}$ molecules and by the $A$ or $B$ irreducible representations of the symmetry group $S_2$ for $C_{2v}$ molecules. The irreducible representation $\Gamma_{ns}$ of a particular nuclear spin state determines its statistical weight and is related to the total nuclear spin $\vec I$ of the state. Here, $\vec I$ is the vector sum of spins $\vec i$ of identical nuclei. 

For H$_3^+$, the $\Gamma_{ns}=A_1$ states ($A_2'$ and $A_2''$ rovibrational states) correspond to $I=3/2$ (ortho); the $\Gamma_{ns}=E$ states ($E'$ and $E''$ rovibrational states) correspond to $I=1/2$ (para). The statistical ortho:para weights are $2:1$. 

For H$_2$D$^+$, the $\Gamma_{ns}=A$ states ($B_1$ and $B_2$ rovibrational states) correspond to $I=1$ (ortho); the $\Gamma_{ns}=B$ states ($A_1$ and $A_2$ rovibrational states) correspond to $I=0$ (para). The statistical ortho:para weights are $3:1$.  

For D$_2$H$^+$, the $\Gamma_{ns}=A$ states ($A_1$ and $A_2$ rovibrational states) correspond to $I=0,2$ (ortho); the $\Gamma_{ns}=B$ states ($B_1$ and $B_2$ rovibrational states) correspond to $I=1$ (para). The statistical ortho:para weights are $2:1$. 

Finally, for D$_3^+$, the $\Gamma_{ns}=A_1$ states ($A_1'$ and $A_1''$ rovibrational states) correspond to $I=1,3$ (ortho); the $\Gamma_{ns}=A_2$ states ($A_2'$ and $A_2''$ rovibrational states) correspond to $I=0$ (para); the $\Gamma_{ns}=E$ states ($E'$ and $E''$ rovibrational states) correspond to $I=1,2$ (meta). The statistical ortho:para:meta weights are $10:1:8$. 

\begin{figure}
\includegraphics[width=8cm]{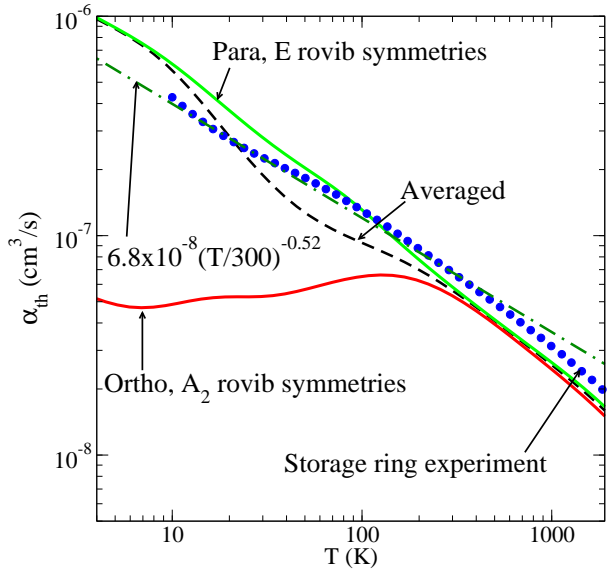}
\caption{Theoretical DR rate coefficients as functions of temperature for the ortho-- and para--species of H$_3^+$. The figure also shows the species-averaged rate coefficient. For comparison, we show the rate coefficient obtained from measurements in the TSR storage ring by \citet{kreckel05} and the analytical dependence for the coefficient used in earlier models of prestellar cores by FPdFW.}
\label{fig:thermal-rate_1}
\end{figure}

\begin{figure}
\includegraphics[width=8cm]{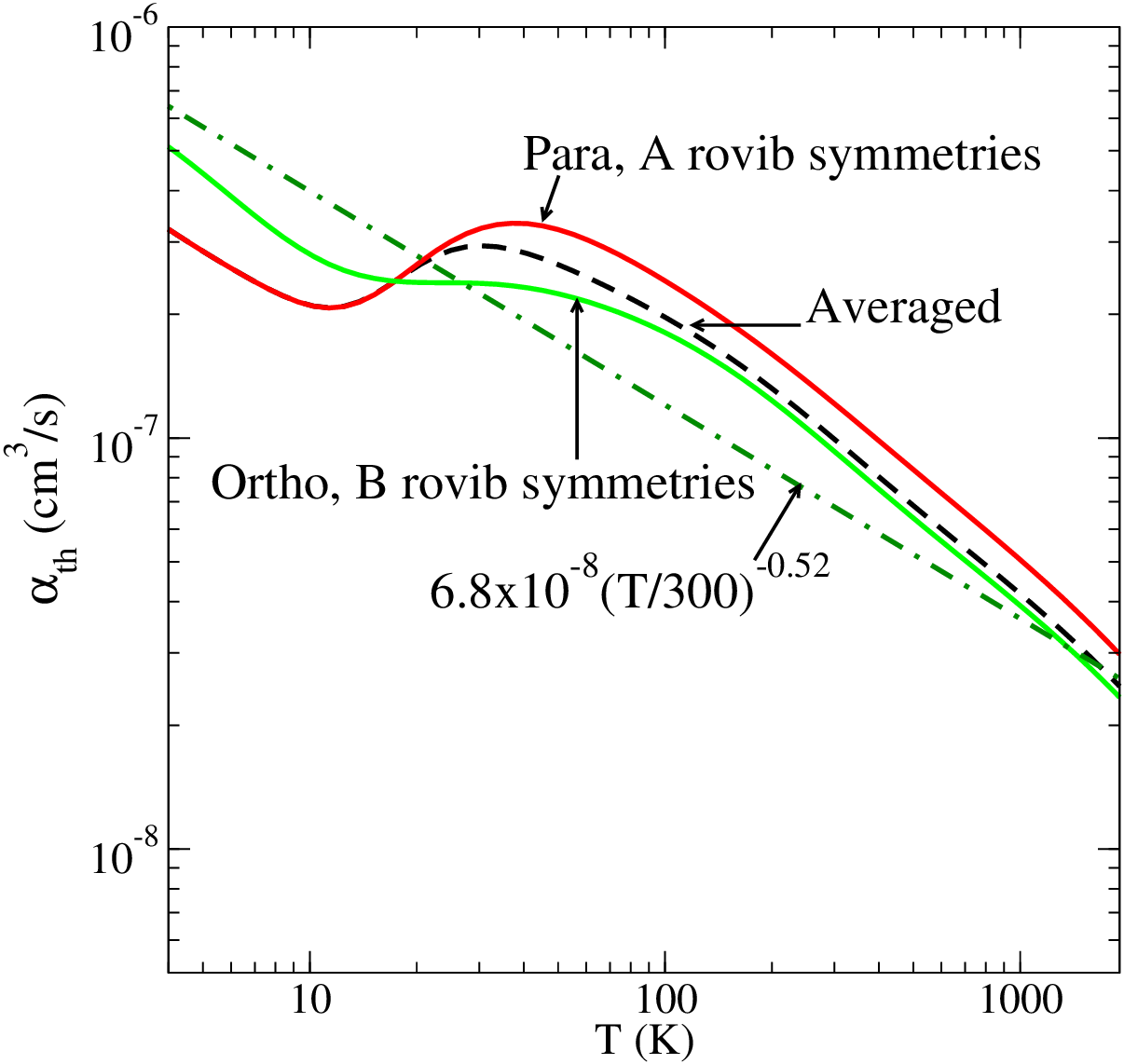}
\caption{Theoretical DR rate coefficients for the ortho-- and para--species of H$_2$D$^+$, as functions of temperature. The rate coefficient averaged over the two species and the analytical expression used in earlier models are also shown.}
\label{fig:thermal-rate_2}
\end{figure} 

\begin{figure}
\includegraphics[width=8cm]{0587fg12.eps}
\caption{Theoretical DR rate coefficients for the ortho-- and para--species of D$_2$H$^+$ as a function of temperature. The rate coefficient averaged over the two species and the analytical expression used in earlier models are also shown.}
\label{fig:thermal-rate_3}
\end{figure}

\begin{figure}
\includegraphics[width=8cm]{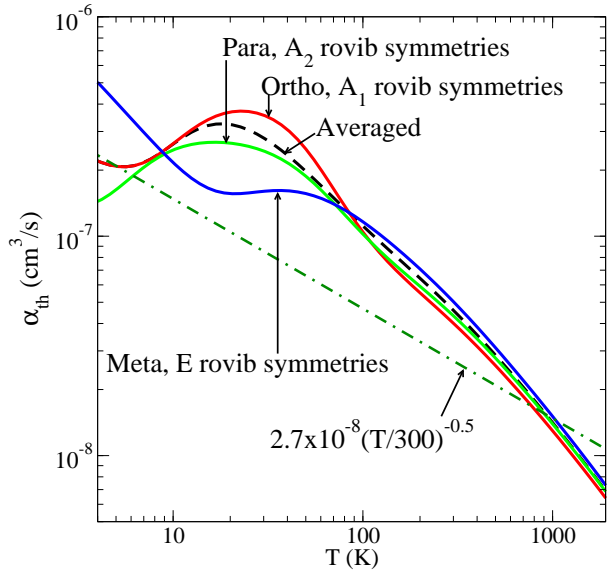}
\caption{Theoretical DR rate coefficients for the ortho--, para--, and meta-species of D$_3^+$. The rate coefficient averaged over the two species and the one used in earlier models are also shown.}
\label{fig:thermal-rate_4}
\end{figure} 

Figures (\ref{fig:thermal-rate_1}), (\ref{fig:thermal-rate_2}), (\ref{fig:thermal-rate_3}), and (\ref{fig:thermal-rate_4}) summarize the obtained DR thermal rate coefficients calculated separately for each nuclear spin species of the four H$_3^+$ isotopologues and the numerical values are listed in Table \ref{DR}. 
For comparison, the figures also show the analytical dependences used in previous models of prestellar core chemistry (FPdFW). As one can see, the rates for different nuclear spin species are similar to each other (for a given isotopologue) at high temperatures. However, for lower temperatures, the rates for different ortho/para/meta-nuclear spin species significantly differ from each other. The difference in behavior at low temperatures is explained by different energies of Rydberg resonances present in DR cross-sections at low electron energies. The actual energies of such resonances are important for the thermal average at temperatures below or similar to the energy difference between ground rotational levels of different nuclear spin species. At higher temperatures, the exact energy of the resonances is not important. The averaged rate is determined by the density and the widths of the resonances, which are similar for all nuclear spin species over a large range of collision energies.

\onecolumn
\begin{longtable}
{rccccccccc}
%\caption{Reactions used in the model. The rate coefficient is given at 7K.\label{reactions}}
\hline
%Temperature &  para--\hddp  &  ortho--\hddp &   ortho--\ddhp  &  para--\ddhp &  ortho--\dtp  &  meta--\dtp  & para--\dtp \\
%             K          &  cm$^3$~s$^{-1}$&  cm$^3$~s$^{-1}$&  cm$^3$~s$^{-1}$&  cm$^3$~s$^{-1}$&  cm$^3$~s$^{-1}$ &  cm$^3$~s$^{-1}$&  cm$^3$~s$^{-1}$ \\
Temperature&para--\htp&ortho--\htp  &  para--\hddp  &  ortho--\hddp &   ortho--\ddhp  &  para--\ddhp &  ortho--\dtp  &  meta--\dtp  & para--\dtp \\
             K          &  cm$^3$~s$^{-1}$&  cm$^3$~s$^{-1}$&  cm$^3$~s$^{-1}$&  cm$^3$~s$^{-1}$&  cm$^3$~s$^{-1}$&  cm$^3$~s$^{-1}$&  cm$^3$~s$^{-1}$ &  cm$^3$~s$^{-1}$&  cm$^3$~s$^{-1}$ \\
\hline
\endhead
\endfoot
    1.28 & 1.81e-06 & 8.21e-08 & 6.63e-07 & 8.74e-07 & 2.18e-07 & 9.59e-08 & 6.33e-07 & 2.38e-06 & 2.61e-07\\
    1.41 & 1.71e-06 & 7.86e-08 & 6.23e-07 & 8.46e-07 & 2.02e-07 & 9.36e-08 & 5.90e-07 & 2.22e-06 & 2.47e-07\\
    1.56 & 1.61e-06 & 7.52e-08 & 5.85e-07 & 8.17e-07 & 1.87e-07 & 9.16e-08 & 5.49e-07 & 2.06e-06 & 2.34e-07\\
    1.72 & 1.52e-06 & 7.20e-08 & 5.48e-07 & 7.87e-07 & 1.72e-07 & 8.98e-08 & 5.10e-07 & 1.90e-06 & 2.22e-07\\
    1.90 & 1.44e-06 & 6.90e-08 & 5.14e-07 & 7.56e-07 & 1.58e-07 & 8.82e-08 & 4.73e-07 & 1.75e-06 & 2.10e-07\\
    2.10 & 1.36e-06 & 6.61e-08 & 4.81e-07 & 7.25e-07 & 1.45e-07 & 8.65e-08 & 4.39e-07 & 1.61e-06 & 1.98e-07\\
    2.32 & 1.29e-06 & 6.34e-08 & 4.51e-07 & 6.93e-07 & 1.32e-07 & 8.47e-08 & 4.08e-07 & 1.48e-06 & 1.88e-07\\
    2.56 & 1.22e-06 & 6.08e-08 & 4.23e-07 & 6.60e-07 & 1.21e-07 & 8.28e-08 & 3.79e-07 & 1.35e-06 & 1.79e-07\\
    2.83 & 1.16e-06 & 5.84e-08 & 3.97e-07 & 6.27e-07 & 1.11e-07 & 8.08e-08 & 3.52e-07 & 1.23e-06 & 1.72e-07\\
    3.12 & 1.11e-06 & 5.62e-08 & 3.73e-07 & 5.94e-07 & 1.02e-07 & 7.87e-08 & 3.29e-07 & 1.12e-06 & 1.66e-07\\
    3.45 & 1.06e-06 & 5.42e-08 & 3.51e-07 & 5.60e-07 & 9.35e-08 & 7.64e-08 & 3.08e-07 & 1.01e-06 & 1.62e-07\\
    3.81 & 1.01e-06 & 5.23e-08 & 3.31e-07 & 5.28e-07 & 8.62e-08 & 7.41e-08 & 2.90e-07 & 9.18e-07 & 1.61e-07\\
    4.20 & 9.63e-07 & 5.08e-08 & 3.13e-07 & 4.95e-07 & 7.98e-08 & 7.17e-08 & 2.74e-07 & 8.30e-07 & 1.63e-07\\
    4.64 & 9.19e-07 & 4.94e-08 & 2.96e-07 & 4.63e-07 & 7.41e-08 & 6.94e-08 & 2.62e-07 & 7.49e-07 & 1.66e-07\\
    5.13 & 8.77e-07 & 4.83e-08 & 2.81e-07 & 4.33e-07 & 6.90e-08 & 6.70e-08 & 2.52e-07 & 6.75e-07 & 1.73e-07\\
    5.66 & 8.36e-07 & 4.76e-08 & 2.67e-07 & 4.04e-07 & 6.45e-08 & 6.47e-08 & 2.46e-07 & 6.07e-07 & 1.80e-07\\
    6.25 & 7.95e-07 & 4.71e-08 & 2.54e-07 & 3.76e-07 & 6.04e-08 & 6.24e-08 & 2.43e-07 & 5.46e-07 & 1.90e-07\\
    6.90 & 7.55e-07 & 4.69e-08 & 2.42e-07 & 3.51e-07 & 5.69e-08 & 6.02e-08 & 2.45e-07 & 4.92e-07 & 1.99e-07\\
    7.62 & 7.14e-07 & 4.71e-08 & 2.31e-07 & 3.28e-07 & 5.37e-08 & 5.81e-08 & 2.50e-07 & 4.42e-07 & 2.09e-07\\
    8.41 & 6.74e-07 & 4.75e-08 & 2.22e-07 & 3.08e-07 & 5.08e-08 & 5.61e-08 & 2.59e-07 & 3.98e-07 & 2.18e-07\\
    9.29 & 6.35e-07 & 4.81e-08 & 2.14e-07 & 2.91e-07 & 4.82e-08 & 5.43e-08 & 2.72e-07 & 3.59e-07 & 2.27e-07\\
   10.25 & 5.96e-07 & 4.88e-08 & 2.09e-07 & 2.77e-07 & 4.59e-08 & 5.28e-08 & 2.87e-07 & 3.25e-07 & 2.34e-07\\
   11.32 & 5.58e-07 & 4.96e-08 & 2.07e-07 & 2.65e-07 & 4.38e-08 & 5.15e-08 & 3.04e-07 & 2.95e-07 & 2.39e-07\\
   12.50 & 5.21e-07 & 5.05e-08 & 2.09e-07 & 2.56e-07 & 4.19e-08 & 5.06e-08 & 3.22e-07 & 2.69e-07 & 2.44e-07\\
   13.80 & 4.86e-07 & 5.12e-08 & 2.14e-07 & 2.49e-07 & 4.02e-08 & 5.01e-08 & 3.39e-07 & 2.48e-07 & 2.46e-07\\
   15.24 & 4.53e-07 & 5.18e-08 & 2.24e-07 & 2.44e-07 & 3.87e-08 & 5.01e-08 & 3.55e-07 & 2.30e-07 & 2.48e-07\\
   16.82 & 4.21e-07 & 5.22e-08 & 2.36e-07 & 2.41e-07 & 3.74e-08 & 5.05e-08 & 3.68e-07 & 2.16e-07 & 2.49e-07\\
   18.57 & 3.92e-07 & 5.25e-08 & 2.52e-07 & 2.40e-07 & 3.65e-08 & 5.15e-08 & 3.79e-07 & 2.06e-07 & 2.48e-07\\
   20.51 & 3.65e-07 & 5.26e-08 & 2.68e-07 & 2.39e-07 & 3.58e-08 & 5.30e-08 & 3.85e-07 & 1.98e-07 & 2.46e-07\\
   22.64 & 3.40e-07 & 5.26e-08 & 2.85e-07 & 2.39e-07 & 3.56e-08 & 5.49e-08 & 3.87e-07 & 1.92e-07 & 2.43e-07\\
   25.00 & 3.17e-07 & 5.26e-08 & 3.01e-07 & 2.39e-07 & 3.58e-08 & 5.73e-08 & 3.85e-07 & 1.89e-07 & 2.40e-07\\
   27.60 & 2.96e-07 & 5.27e-08 & 3.14e-07 & 2.38e-07 & 3.64e-08 & 5.98e-08 & 3.79e-07 & 1.86e-07 & 2.35e-07\\
   30.48 & 2.77e-07 & 5.29e-08 & 3.24e-07 & 2.38e-07 & 3.75e-08 & 6.26e-08 & 3.68e-07 & 1.83e-07 & 2.28e-07\\
   33.65 & 2.59e-07 & 5.32e-08 & 3.31e-07 & 2.37e-07 & 3.92e-08 & 6.53e-08 & 3.53e-07 & 1.81e-07 & 2.21e-07\\
   37.15 & 2.44e-07 & 5.38e-08 & 3.33e-07 & 2.35e-07 & 4.14e-08 & 6.78e-08 & 3.34e-07 & 1.78e-07 & 2.12e-07\\
   41.02 & 2.29e-07 & 5.46e-08 & 3.32e-07 & 2.32e-07 & 4.42e-08 & 7.01e-08 & 3.11e-07 & 1.75e-07 & 2.02e-07\\
   45.29 & 2.16e-07 & 5.57e-08 & 3.28e-07 & 2.29e-07 & 4.76e-08 & 7.21e-08 & 2.85e-07 & 1.71e-07 & 1.91e-07\\
   50.01 & 2.04e-07 & 5.68e-08 & 3.22e-07 & 2.24e-07 & 5.15e-08 & 7.36e-08 & 2.57e-07 & 1.66e-07 & 1.79e-07\\
   55.21 & 1.92e-07 & 5.81e-08 & 3.13e-07 & 2.20e-07 & 5.59e-08 & 7.48e-08 & 2.29e-07 & 1.60e-07 & 1.67e-07\\
   60.96 & 1.81e-07 & 5.95e-08 & 3.03e-07 & 2.14e-07 & 6.06e-08 & 7.56e-08 & 2.01e-07 & 1.54e-07 & 1.54e-07\\
   67.31 & 1.71e-07 & 6.08e-08 & 2.91e-07 & 2.08e-07 & 6.54e-08 & 7.61e-08 & 1.76e-07 & 1.47e-07 & 1.42e-07\\
   74.31 & 1.61e-07 & 6.21e-08 & 2.79e-07 & 2.02e-07 & 7.01e-08 & 7.63e-08 & 1.54e-07 & 1.40e-07 & 1.30e-07\\
   82.05 & 1.51e-07 & 6.33e-08 & 2.67e-07 & 1.95e-07 & 7.44e-08 & 7.63e-08 & 1.34e-07 & 1.33e-07 & 1.19e-07\\
   90.59 & 1.41e-07 & 6.43e-08 & 2.54e-07 & 1.88e-07 & 7.82e-08 & 7.61e-08 & 1.18e-07 & 1.25e-07 & 1.09e-07\\
  100.02 & 1.31e-07 & 6.52e-08 & 2.41e-07 & 1.81e-07 & 8.12e-08 & 7.56e-08 & 1.05e-07 & 1.18e-07 & 1.00e-07\\
  110.43 & 1.22e-07 & 6.58e-08 & 2.29e-07 & 1.73e-07 & 8.33e-08 & 7.50e-08 & 9.35e-08 & 1.10e-07 & 9.21e-08\\
  121.93 & 1.13e-07 & 6.61e-08 & 2.17e-07 & 1.65e-07 & 8.45e-08 & 7.42e-08 & 8.42e-08 & 1.03e-07 & 8.50e-08\\
  134.62 & 1.05e-07 & 6.61e-08 & 2.05e-07 & 1.57e-07 & 8.46e-08 & 7.31e-08 & 7.65e-08 & 9.58e-08 & 7.87e-08\\
  148.64 & 9.70e-08 & 6.58e-08 & 1.93e-07 & 1.48e-07 & 8.40e-08 & 7.19e-08 & 7.00e-08 & 8.91e-08 & 7.31e-08\\
  164.11 & 8.99e-08 & 6.50e-08 & 1.82e-07 & 1.40e-07 & 8.25e-08 & 7.04e-08 & 6.44e-08 & 8.27e-08 & 6.80e-08\\
  181.20 & 8.35e-08 & 6.38e-08 & 1.71e-07 & 1.32e-07 & 8.05e-08 & 6.86e-08 & 5.94e-08 & 7.67e-08 & 6.34e-08\\
  200.06 & 7.76e-08 & 6.23e-08 & 1.60e-07 & 1.23e-07 & 7.80e-08 & 6.67e-08 & 5.50e-08 & 7.09e-08 & 5.90e-08\\
  220.89 & 7.23e-08 & 6.04e-08 & 1.50e-07 & 1.15e-07 & 7.51e-08 & 6.46e-08 & 5.11e-08 & 6.54e-08 & 5.49e-08\\
  243.88 & 6.73e-08 & 5.83e-08 & 1.40e-07 & 1.08e-07 & 7.21e-08 & 6.23e-08 & 4.73e-08 & 6.02e-08 & 5.10e-08\\
  269.27 & 6.28e-08 & 5.59e-08 & 1.31e-07 & 1.00e-07 & 6.90e-08 & 5.99e-08 & 4.38e-08 & 5.53e-08 & 4.73e-08\\
  297.31 & 5.87e-08 & 5.34e-08 & 1.22e-07 & 9.34e-08 & 6.58e-08 & 5.74e-08 & 4.04e-08 & 5.07e-08 & 4.37e-08\\
  328.26 & 5.48e-08 & 5.08e-08 & 1.14e-07 & 8.68e-08 & 6.27e-08 & 5.49e-08 & 3.73e-08 & 4.64e-08 & 4.03e-08\\
  362.43 & 5.13e-08 & 4.81e-08 & 1.06e-07 & 8.07e-08 & 5.97e-08 & 5.24e-08 & 3.43e-08 & 4.23e-08 & 3.71e-08\\
  400.16 & 4.80e-08 & 4.54e-08 & 9.85e-08 & 7.51e-08 & 5.67e-08 & 4.99e-08 & 3.15e-08 & 3.86e-08 & 3.41e-08\\
  441.82 & 4.49e-08 & 4.27e-08 & 9.17e-08 & 6.99e-08 & 5.38e-08 & 4.74e-08 & 2.88e-08 & 3.51e-08 & 3.12e-08\\
  487.81 & 4.21e-08 & 4.01e-08 & 8.53e-08 & 6.50e-08 & 5.11e-08 & 4.51e-08 & 2.63e-08 & 3.18e-08 & 2.85e-08\\
  538.60 & 3.95e-08 & 3.76e-08 & 7.94e-08 & 6.06e-08 & 4.85e-08 & 4.27e-08 & 2.40e-08 & 2.88e-08 & 2.60e-08\\
  594.67 & 3.70e-08 & 3.52e-08 & 7.40e-08 & 5.65e-08 & 4.59e-08 & 4.05e-08 & 2.19e-08 & 2.61e-08 & 2.36e-08\\
  656.58 & 3.47e-08 & 3.29e-08 & 6.88e-08 & 5.27e-08 & 4.35e-08 & 3.83e-08 & 1.99e-08 & 2.35e-08 & 2.14e-08\\
  724.93 & 3.26e-08 & 3.08e-08 & 6.41e-08 & 4.91e-08 & 4.11e-08 & 3.62e-08 & 1.80e-08 & 2.12e-08 & 1.94e-08\\
  800.40 & 3.05e-08 & 2.87e-08 & 5.96e-08 & 4.58e-08 & 3.88e-08 & 3.41e-08 & 1.63e-08 & 1.91e-08 & 1.76e-08\\
  883.72 & 2.86e-08 & 2.68e-08 & 5.55e-08 & 4.27e-08 & 3.66e-08 & 3.20e-08 & 1.47e-08 & 1.72e-08 & 1.59e-08\\
  975.72 & 2.68e-08 & 2.50e-08 & 5.15e-08 & 3.98e-08 & 3.43e-08 & 3.01e-08 & 1.33e-08 & 1.55e-08 & 1.43e-08\\
 1077.30 & 2.51e-08 & 2.33e-08 & 4.78e-08 & 3.70e-08 & 3.22e-08 & 2.81e-08 & 1.20e-08 & 1.39e-08 & 1.29e-08\\
 1189.45 & 2.35e-08 & 2.17e-08 & 4.43e-08 & 3.44e-08 & 3.00e-08 & 2.62e-08 & 1.08e-08 & 1.25e-08 & 1.16e-08\\
 1313.28 & 2.20e-08 & 2.02e-08 & 4.09e-08 & 3.19e-08 & 2.80e-08 & 2.43e-08 & 9.70e-09 & 1.12e-08 & 1.04e-08\\
 1450.00 & 2.05e-08 & 1.87e-08 & 3.77e-08 & 2.95e-08 & 2.59e-08 & 2.25e-08 & 8.70e-09 & 1.00e-08 & 9.33e-09\\
 1600.95 & 1.90e-08 & 1.73e-08 & 3.47e-08 & 2.72e-08 & 2.40e-08 & 2.08e-08 & 7.80e-09 & 8.95e-09 & 8.36e-09\\
 1767.62 & 1.76e-08 & 1.60e-08 & 3.18e-08 & 2.50e-08 & 2.21e-08 & 1.91e-08 & 6.97e-09 & 7.99e-09 & 7.47e-09\\
 1951.64 & 1.63e-08 & 1.47e-08 & 2.91e-08 & 2.29e-08 & 2.02e-08 & 1.75e-08 & 6.23e-09 & 7.12e-09 & 6.67e-09\\
 2154.81 & 1.50e-08 & 1.35e-08 & 2.66e-08 & 2.09e-08 & 1.85e-08 & 1.60e-08 & 5.55e-09 & 6.34e-09 & 5.94e-09\\
 2379.14 & 1.38e-08 & 1.24e-08 & 2.42e-08 & 1.91e-08 & 1.69e-08 & 1.45e-08 & 4.94e-09 & 5.64e-09 & 5.29e-09\\
 2626.82 & 1.26e-08 & 1.13e-08 & 2.19e-08 & 1.73e-08 & 1.53e-08 & 1.32e-08 & 4.39e-09 & 5.00e-09 & 4.70e-09\\
 2900.28 & 1.15e-08 & 1.03e-08 & 1.98e-08 & 1.57e-08 & 1.38e-08 & 1.19e-08 & 3.90e-09 & 4.44e-09 & 4.17e-09\\
 3202.22 & 1.04e-08 & 9.36e-09 & 1.79e-08 & 1.42e-08 & 1.25e-08 & 1.07e-08 & 3.46e-09 & 3.93e-09 & 3.70e-09\\
 3535.58 & 9.48e-09 & 8.47e-09 & 1.61e-08 & 1.28e-08 & 1.12e-08 & 9.65e-09 & 3.06e-09 & 3.48e-09 & 3.27e-09\\
 3903.65 & 8.57e-09 & 7.66e-09 & 1.45e-08 & 1.15e-08 & 1.01e-08 & 8.66e-09 & 2.71e-09 & 3.08e-09 & 2.90e-09\\
 4310.04 & 7.75e-09 & 6.91e-09 & 1.30e-08 & 1.03e-08 & 9.04e-09 & 7.76e-09 & 2.40e-09 & 2.72e-09 & 2.57e-09\\
 4758.74 & 6.99e-09 & 6.23e-09 & 1.16e-08 & 9.26e-09 & 8.10e-09 & 6.95e-09 & 2.13e-09 & 2.41e-09 & 2.27e-09\\
 5254.15 & 6.31e-09 & 5.61e-09 & 1.04e-08 & 8.31e-09 & 7.26e-09 & 6.23e-09 & 1.89e-09 & 2.14e-09 & 2.02e-09\\
 5801.13 & 5.70e-09 & 5.06e-09 & 9.39e-09 & 7.47e-09 & 6.51e-09 & 5.58e-09 & 1.68e-09 & 1.90e-09 & 1.79e-09\\
 6405.05 & 5.15e-09 & 4.58e-09 & 8.45e-09 & 6.73e-09 & 5.85e-09 & 5.02e-09 & 1.50e-09 & 1.69e-09 & 1.60e-09\\
 7071.85 & 4.67e-09 & 4.14e-09 & 7.63e-09 & 6.08e-09 & 5.27e-09 & 4.52e-09 & 1.34e-09 & 1.51e-09 & 1.43e-09\\
 7808.06 & 4.24e-09 & 3.77e-09 & 6.91e-09 & 5.50e-09 & 4.77e-09 & 4.08e-09 & 1.20e-09 & 1.36e-09 & 1.29e-09\\
 8620.92 & 3.87e-09 & 3.43e-09 & 6.28e-09 & 5.01e-09 & 4.33e-09 & 3.71e-09 & 1.09e-09 & 1.23e-09 & 1.16e-09\\
 9518.40 & 3.55e-09 & 3.14e-09 & 5.73e-09 & 4.57e-09 & 3.95e-09 & 3.38e-09 & 9.86e-10 & 1.11e-09 & 1.05e-09\\
10509.31 & 3.26e-09 & 2.89e-09 & 5.26e-09 & 4.20e-09 & 3.61e-09 & 3.09e-09 & 8.99e-10 & 1.01e-09 & 9.60e-10\\
11603.38 & 3.01e-09 & 2.67e-09 & 4.85e-09 & 3.87e-09 & 3.33e-09 & 2.85e-09 & 8.25e-10 & 9.29e-10 & 8.80e-10\\
12811.35 & 2.80e-09 & 2.48e-09 & 4.50e-09 & 3.58e-09 & 3.08e-09 & 2.63e-09 & 7.61e-10 & 8.56e-10 & 8.12e-10\\
14145.07 & 2.61e-09 & 2.31e-09 & 4.19e-09 & 3.34e-09 & 2.86e-09 & 2.45e-09 & 7.05e-10 & 7.94e-10 & 7.53e-10\\
15617.64 & 2.45e-09 & 2.16e-09 & 3.92e-09 & 3.12e-09 & 2.67e-09 & 2.29e-09 & 6.58e-10 & 7.40e-10 & 7.01e-10\\
17243.51 & 2.30e-09 & 2.04e-09 & 3.68e-09 & 2.94e-09 & 2.51e-09 & 2.14e-09 & 6.16e-10 & 6.93e-10 & 6.57e-10\\
19038.65 & 2.18e-09 & 1.92e-09 & 3.48e-09 & 2.77e-09 & 2.37e-09 & 2.02e-09 & 5.80e-10 & 6.52e-10 & 6.19e-10\\
21020.66 & 2.07e-09 & 1.83e-09 & 3.30e-09 & 2.63e-09 & 2.24e-09 & 1.92e-09 & 5.49e-10 & 6.17e-10 & 5.85e-10\\
23209.01 & 1.97e-09 & 1.74e-09 & 3.15e-09 & 2.51e-09 & 2.13e-09 & 1.82e-09 & 5.22e-10 & 5.86e-10 & 5.56e-10\\
\hline
\caption{Dissociative recombination rates of \htp, \hddp, \ddhp, and \dtpb for each individual nuclear spin state species.\label{DR}}

\end{longtable}
\end{appendix}

\end{document}